\documentclass[twocolumn]{revtex4}
\usepackage{graphicx}
\newcommand {\phidot}{\dot{\phi}} 
\newcommand {\phitwo}{\ddot{\phi}}

\newcommand {\mpl}{m_{\scriptscriptstyle{PL}}}

\def\cmm2{{\,\rm cm^{-2}}}
\def\cm2{{\,{\rm cm}^2}}
\def\cmm3{{\,{\rm cm}^{-3}}}
\def\gcmm3{{\,{\rm g\,cm^{-3}}}}

\def\mpl{{m_{\rm Pl}}}

\def\la{\mathrel{\mathpalette\fun <}}
\def\ga{\mathrel{\mathpalette\fun >}}
\def\fun#1#2{\lower3.6pt\vbox{\baselineskip0pt\lineskip.9pt
  \ialign{$\mathsurround=0pt#1\hfil##\hfil$\crcr#2\crcr\sim\crcr}}}

\begin{document}

\title{Probing the dark energy: methods and strategies}
\author{Dragan Huterer$^1$ and Michael S. Turner$^{1,2,3}$
\vspace{0.15cm}}

\affiliation{\it $^1$Department of Physics\\ Enrico Fermi
Institute \\ The University of Chicago \\
Chicago, IL~~60637-1433\vspace{0.15cm}} 

\affiliation{\it 
$^2$Department of Astronomy \& Astrophysics \\
The University of Chicago\\  Chicago, IL~~60637-1433\vspace{0.15cm}}

\affiliation{\it $^3$NASA/Fermilab Astrophysics Center\\
Fermi National Accelerator Laboratory \\ Batavia, IL~~60510-0500}

\begin{abstract}
The presence of dark energy in the Universe is inferred directly
from the accelerated expansion of the Universe, and indirectly,
from measurements of cosmic microwave background (CMB)
anisotropy.  Dark energy contributes about 2/3 of the critical
density, is very smoothly distributed, and has large negative
pressure.  Its nature is very much unknown.  Most of its
discernible consequences follow from its effect on evolution of
the expansion rate of the Universe, which in turn affects the
growth of density perturbations and the age of the Universe, and
can be probed by the classical kinematic cosmological tests. Absent a
compelling theoretical model (or even a class of models), we
describe the dark energy by an effective equation-of-state
$w=p_X/\rho_X$ which is allowed to vary with time.   We
describe and compare different approaches for determining $w(t)$,
including magnitude-redshift (Hubble) diagram, number counts of
galaxies and clusters, and CMB anisotropy, focusing particular
attention on the use of a sample of several thousand type Ia
supernova with redshifts $z\lesssim 1.7$, as might be gathered by
the proposed SNAP satellite. Among other things, we derive
optimal strategies for constraining cosmological parameters using
type Ia supernovae. While in the near term CMB anisotropy will
provide the first measurements of $w$, supernovae and number
counts appear to have the most potential to probe dark energy. 

\end{abstract}

\maketitle

\section{Introduction}

There is good evidence that a mysterious form of dark energy
accounts for about 2/3rds of the matter and energy in the
Universe.  The direct evidence comes from distance measurements
of type Ia supernovae (SNe Ia) which indicate the expansion of
the Universe is speeding up, not slowing down
\cite{perlmutter-1999,riess}.  Equally strong indirect evidence now comes
from the factor of three discrepancy \cite{mst_SNe_review,
dod_knox} between cosmic microwave background (CMB) anisotropy
measurements which indicate $\Omega_0\simeq 1.1\pm 0.07$
\cite{boom,max,Jaffe} and measurements of the matter density
$\Omega_M =0.35\pm 0.07$~\cite{mst_scripta} together with the consistency
between the level of inhomogeneity revealed by CMB anisotropy
and the structure that exists today ($\Omega_0$ is
the fraction of critical density contributed by all forms of matter and energy).
The former implies the existence of a smooth component of energy (or matter)
that contributes 2/3rds of the critical density;
and the latter argues for it having large, negative pressure, which leads
to its repulsive gravity.  Because a smooth component of matter
or energy interferes with the growth of linear density perturbations 
and the formation of structure, the energy density of the smooth component
must evolve more slowly than that of matter.  The amount of growth
needed to form the structure seen today from the initial inhomogeneity
revealed by the CMB implies that the bulk pressure of the smooth
component must be more negative than about $-\rho /2$ \cite{believe}.
(Because its pressure is comparable in magnitude to its energy
density, it is relativistic and energy like -- hence the term
dark energy.)

Finally, additional indirect evidence for dark energy comes from
detailed studies of how galaxies and clusters of
galaxies formed from primeval density perturbations.
The cold dark matter (CDM) paradigm for structure formation
successfully accounts for most of the features of the Universe
we observe today (so much so that there is virtually no competing
theory).  Of the flat CDM models (hot + cold, tilted, enhanced
radiation, or very low Hubble constant) the one with a cosmological 
constant ($\Lambda$CDM) is the most successful and consistent with
virtually all observations \cite{LCDM}.

Even before the evidence for dark energy discussed above, there was a
dark-energy candidate:  the energy density of the quantum vacuum (or
cosmological constant) for which $p=-\rho$.  However,
the inability of particle theorists to compute the energy of the
quantum vacuum -- contributions from well understood physics
amount to $10^{55}$ times critical density -- casts a dark shadow
on the cosmological constant \cite{weinberg_rmp}.  It is possible
that contributions from ``new physics'' add together to nearly
cancel those from known physics, leaving a tiny cosmological
constant.  However, the fine tuning required (a precision
of at least 54 decimal places) makes a complete cancellation
seem more plausible.

If the cosmological constant is zero, something else must be
causing the Universe to speed up.  A host of
other possibilities have now been discussed: rolling scalar field (or
quintessence)~\cite{scalar, zlatev,manojetal,aaetal,k-essence};
a network of frustrated
topological defects~\cite{defect}; the energy of a metastable
vacuum state~\cite{wilczek}; effects having to do with extra
dimensions~\cite{extra}; quantum effects of a massive scalar
field~\cite{parker_raval}; particles with a time-varying mass
\cite{carroll_anderson}; and ``solid'' or ``generalized'' dark
matter~\cite{bucher,gdm}.  While all of these models have some
motivation and attractive features, none are compelling.  On
the other hand, the cosmological constant is extremely well
motivated, but equally problematic.  This in essence is the dark
energy problem.

The two most conspicuous features of dark energy are smooth
spatial distribution and large negative pressure.  While only
vacuum energy is absolutely uniform in its spatial distribution,
all the other examples of dark energy only clump on the largest
scales at a level that can be neglected for most purposes
\cite{smooth} (more on this in the Summary).  Motivated by this
as well as the absence of compelling theoretical model or
framework for dark energy, Turner and White~\cite{turner_white}
have suggested parameterizing the dark energy by its bulk
equation-of-state: $w \equiv \langle p_X \rangle /\langle \rho_X
\rangle$.  For different dark energy models $w$ takes on
different values (e.g., $-1$ for vacuum energy, or $-N/3$ for
topological defects of dimensionality $N$); $w$ can be
time-varying (e.g., in models with a rolling scalar field).  In
this language, the first step toward solving the dark-energy
problem is determining $w(t)$.

While the dark-energy problem involves both cosmology and
fundamental physics, because of its diffuse nature it seems
likely that cosmological rather than laboratory measurements have
the most probative power.  (It has been emphasized that if the
dark energy involves a very light scalar field, there will be a
new long-range force that could be probed in the laboratory
\cite{carroll}).  It is the purpose of this paper to lay out the
cosmological consequences of dark energy that allow its nature to
be probed, and to assess their efficacy.  Further, we present in
more detail some of the calculations that appeared in the SNAP
proposal~\cite{SNAP}.  In Sec.~\ref{prelim-sec} we begin with an
overview of the cosmological observables that may be of use as
well as a discussion of their sensitivity to the dark-energy
equation-of-state $w$.  In Sec.~\ref{w-sec} we discuss the
relative merits of different cosmological observations in probing
the average value of $w$.  Sec.~\ref{wz-sec} addresses strategies
for the more difficult problem of probing the possible time
variation of $w$.  Sec.~\ref{optimal-sec} discusses optimal
strategies for determining dark-energy properties.  In the final
Section we summarize our results and end with some general
remarks.  We note that there are other studies of how best to get
at the nature of the dark energy \cite{maor, nugent, weller,
astier}, and where appropriate we compare results.

\section{Preliminaries}\label{prelim-sec}

Although dark energy does not clump significantly, it does
significantly affect the large-scale dynamics of the Universe, including
the age of the Universe, the growth of density perturbations and 
the classic cosmological tests~\cite{charlton_turner}.  
All of the consequences of dark
energy follow from its effect on the expansion rate:
\begin{eqnarray}
\label{eq:H^2}
H^2 & = &  {8\pi G \over 3} (\rho_M + \rho_X) \\
H(z)^2  &  = & H_0^2\left[ \Omega_M(1+z)^3 + \right. \nonumber\\
    && \left. \Omega_X \exp [3\int_0^z\,(1+w(x))d\ln (1+x)] \right] \nonumber
\end{eqnarray}
where $\Omega_M$ ($\Omega_X$) is the fraction of critical density
contributed by matter (dark energy) today, a flat Universe is
assumed, and the dark-energy term in the second equation follows
from integrating its equation of motion, $d(\rho_X a^3) = -p_X
da^3$ ($a$ is the cosmic scale factor).\footnotemark[2]
\footnotetext[2]{We have implicitly
assumed that $w=w(z)$.  In general, this need not be the case.
If, for example, we had assumed
$w=w(\rho)$, then $\rho_X$ could not have been expressed in closed
form. Nevertheless, Eq.\ (\ref{eq:H^2}) can be solved if it is
supplemented by the equation governing the behavior
of $\rho_X$, $d\ln\rho_X/(1+w(\rho_X))=-3d\ln a$. Another
example is a minimally coupled scalar field,
where $\rho_X=\phidot^2/2+V(\phi)$,
and its evolution is determined by the equation of
motion of the scalar field, $\phitwo+3H\phidot+V'(\phi)=0$.}

\subsection{Age and growth of density perturbations}

\begin{figure}
\includegraphics[height=3in, width= 2.5in, angle=-90]{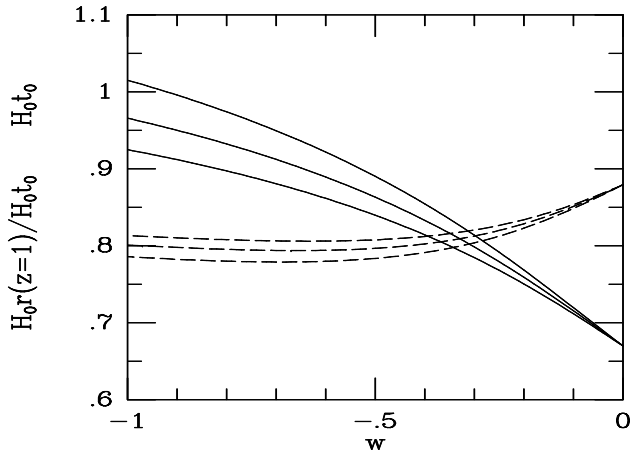}
\caption{Age times Hubble constant as a function of
(constant) $w$ for $\Omega_M = 0.25,0.3,0.35$ (solid
curves, top to bottom); current
measurements indicate that $H_0t_0=0.93\pm 0.15$.
To illustrate the degeneracy between age and comoving
distance measurements, we plot their ratio (broken curves;
top to bottom, $\Omega_M=0.35, 0.30, 0.25$).
Note, this ratio is very insensitive to $\Omega_M$.
}
\label{fig:age}
\end{figure}

The age of the Universe today is related to the expansion history
of the Universe,
\begin{equation}
t_0 = \int_0^{t_0}\,dt = \int_0^\infty {dz\over (1+z)H(z)},
\end{equation}
which depends upon the equation-of-state of the dark energy.  The
more negative $w$ is, the more accelerated the expansion is and
the older the Universe is today for fixed $H_0$ (see
Fig.~\ref{fig:age}).  To make use of this requires accurate
measurements of $H_0$ and $t_0$.  Because the uncertainties in
each are about 10\% (with possible additional systematic errors),
age of the Universe is not an accurate probe of $w$.  In any case,
current measurements, $H_0=70\pm 7\,{\rm km\,sec^{-1}\,
Mpc^{-1}}$ and $t_0=13\pm 1.5\,$Gyr~\cite{age}, imply $H_0t_0
=0.93\pm 0.15$ and favor $w\la -{1/2}$.

The dependences of $H_0t_0$ and $r(z)$ upon $w$ are very similar
for $z\sim 0.5-2$, and further, their ratio is insensitive to
$\Omega_M$ (see Fig.~\ref{fig:age}).  Thus, a measurement of
$H_0t_0$ can add little complementary information to that
provided by precise determinations of $r(z)$.  Of course, because
of this degeneracy, there is a valuable consistency check and
measurements of $r(z)$ have great leverage in fixing $H_0t_0$.
None of this is very surprising since the formulas for $t_0$ and
$r(z)$ are very similar.

\begin{figure}
\includegraphics[height=2.5in, width= 3in]{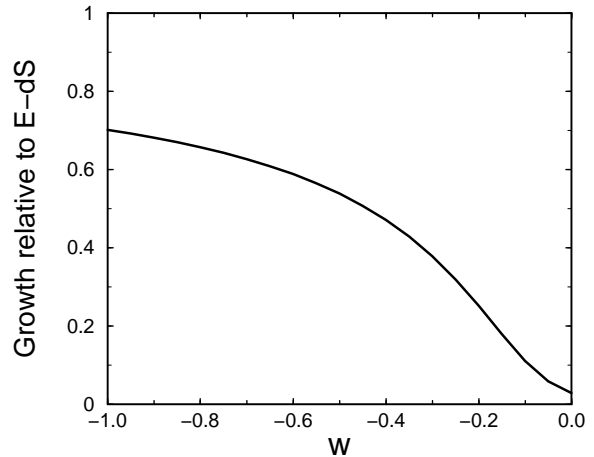}
\caption{Growth of linear perturbations since redshift $z=1000$
relative to the Einstein--deSitter model as a function of
(constant) $w$ for $\Omega_M =0.3$. The dark energy is assumed
not to clump.}
\label{fig:growth}
\end{figure}

The effect on density perturbations is to suppress the growth in
the linear regime, relative to the Einstein-deSitter model, where
the growth is proportional to the cosmic scale factor.  The growth of
linear perturbations is governed by the familiar equation,
\begin{equation}
\ddot\delta_k + 2H\dot\delta_k -4\pi G \rho_M \delta_k = 0
\end{equation}

\noindent where density perturbations in the pressureless cold
dark matter have been decomposed into their Fourier modes, $k$ is
the comoving wavenumber of the mode, and it is assumed that
$k\gg H_0$.  As can be
seen in Fig.~\ref{fig:growth}, the effect on the growth of linear
perturbations is not very significant for $w\la -{1\over 2}$, which
is one of the virtues of dark-energy models since the level of
inhomogeneity revealed in the CMB is just about right to explain
the structure seen today.

The reason the growth is not affected much is because for $w\la
-{1\over 2}$ the Universe only recently became dark-energy
dominated ($\rho_X \ge \rho_M$ for $1+z \le 1+z_X =
(\Omega_X/\Omega_M)^{-1/3w}$), and the growth of perturbations is
essentially the same as in a matter-dominated model until then.
Growth suppression increases with increasing $w$ since the onset
of dark-energy domination occurs earlier (see
Fig.~\ref{fig:growth}).  For $w\ga -{1\over 2}$ the suppression
of the growth of linear perturbations is sufficiently large that
structure observed today could not have evolved from the density
perturbations revealed by CMB
anisotropy~\cite{believe,turner_white}.

\begin{figure}
\includegraphics[height=2.5in, width= 3in]{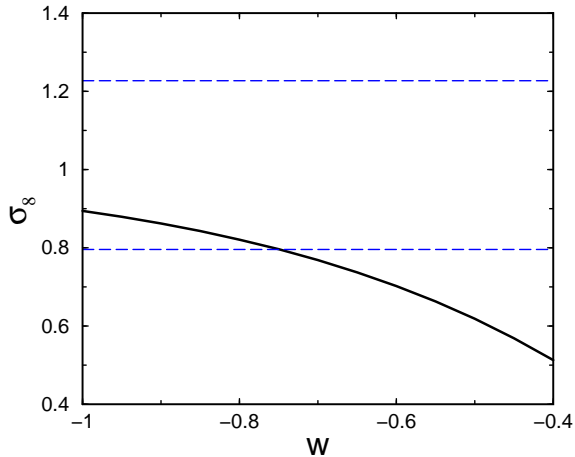}
\caption{The rms amplitude of matter perturbations on the scale
$8h^{-1}\,$Mpc as a function of (constant) $w$ for a COBE
normalized, scale-invariant model with $h=0.7$
(see Ref.~\cite{turner_white} for details).  The present cluster
abundance fixes $\sigma_8 = (0.56\pm 0.1)\Omega_M^{-0.47}$
(95\% cl)~\cite{viana_liddle}, indicated by the broken
lines for $\Omega_M=0.3$.
The downward trend in $\sigma_8$ with increasing $w$ is the
suppression of the growth of linear density perturbations as dark
energy domination occurs earlier, and leads to an upper limit
to $w$ of around $-1/2$.}
\label{fig:sigma_8}
\end{figure}

To be more specific, the suppression of growth affects the overall
normalization of the power spectrum today, most easily expressed
in terms of the {\em rms} mass fluctuations in spheres of
$8h^{-1}\,$Mpc, or $\sigma_8$ (see Fig.~\ref{fig:sigma_8}).
Further, the number density of
bound objects formed by a given redshift is exponentially
sensitive to the growth of density perturbations
\cite{bahcall}. The number density can be
accurately estimated by the Press-Schechter formalism~\cite{PS},
\begin{eqnarray}
\hspace{-0.10cm}
&&\hspace{1cm} \frac{dn}{dM}(z,M)= \nonumber \\
&&\sqrt{\frac{2}{\pi}} \frac{\rho_M}{M}
\frac{\delta_c}{\sigma(M, z)^2}
\frac{d\sigma(M, z)}{dM}\exp\left( -\frac{\delta_c^2}
                                    {2\sigma(M, z)^2}\right)
\label{eq:dndm}
\end{eqnarray}

\noindent where $\sigma(M, z)$ is the rms density fluctuation on
mass-scale $M$ evaluated at redshift $z$ and computed using
linear theory, $\rho_M$ is the present-day matter density, and
$\delta_c\approx 1.68$ is the linear threshold overdensity for
collapse.  

Strong and weak gravitational lensing may be used to constrain
the growth of structure and thus probe the dark energy. We will
not address them here as detailed modeling of the lenses, their
distribution, and the evolution of non-linear structure is
required to address their efficacy.  We refer the reader to
Refs.~\cite{lensing}.  The effect of dark energy on the growth of
linear density perturbations enters in the cluster-number-count
test, as discussed below.

\subsection{Classical tests}

The other cosmological probes of the
dark energy involve the classical tests:
magnitude vs.\ redshift (Hubble) diagram, number-count vs.\
redshift, and angular size vs.\ redshift.  For the flat models
that we consider, all of these depend upon the comoving distance
to an object at redshift $z$, which is determined by the
expansion history:
\begin{equation}
r(z)  =  \int_0^z \,{dx\over H(x)}\,.
\end{equation}

Luminosity distance, which is the distance inferred from measurements of
the apparent luminosity of an object of known intrinsic
luminosity, $\log(d_L(z))
\equiv 0.2 (m - M) - 5$, is related to $r(z)$
\begin{equation}
d_L(z) = (1+z)r(z),
\end{equation}

\noindent where $m$ is apparent luminosity, $M$ the absolute luminosity and
distances are measured in Mpc.  The magnitude -- redshift
(Hubble) diagram is a plot of $m(z)$ vs.\ $z$.

The angular-diameter distance, which is the distance inferred from the
angular size of an object of known size, $d_A(z) = D/\theta$,
is related to $r(z)$
\begin{equation}
d_A = r(z)/(1+z).
\end{equation}
The angular-diameter distance comes into play in using
CMB anisotropy (more below) or the Alcock-Paczynski test
to probe the dark energy.

The Alcock-Paczynski compares the angular size of an object
on the sky with its the redshift extent~\cite{APtest}.
The diameter $D$ of a spherical object (of fixed size
or comoving with the expansion) at redshift $z$
is related to its angular size on the sky $d\theta$ by
$d_A(z)d\theta$ and to its redshift extent
by $\Delta z/(1+z)H(z)$.  Thus, measurements of $\Delta z$ and
$\Delta\theta$ can be combined to determine $H(z)r(z)$:
\begin{equation}
H(z)r(z) = {\Delta z\over \Delta \theta}
\end{equation}

The trick is to find objects (or ensembles of objects) that are
spherical.  One idea involves the correlation function of
galaxies or of Lyman-$\alpha$ clouds, which, because of the
isotropy of the Universe, should have the same dependence upon
separation along the line-of-sight or across the sky.  A large
and uniform sample of objects is needed to implement this test;
further, the effects of peculiar velocities induced by density
perturbations must be separated from the small (5\% or so)
cosmological effect~\cite{ballinger}.

\begin{figure}
\includegraphics[height=2.5in, width= 3in]{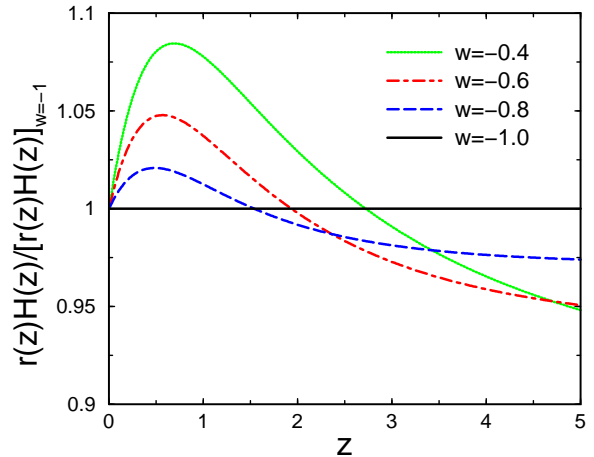}
\caption{The Alcock-Paczynski test, which compares
the angular size ($\Delta \theta$) of a spherical object
at redshift $z$ to its redshift extent ($\Delta z$),
can determine $r(z)H(z)$.  Its sensitivity is shown here
for $\Omega_M=0.3$ and constant $w=-0.4, -0.6, -0.8, -1.0$.
}
\label{fig:AP}
\end{figure}

The authors of Ref.~\cite{lamhuietal} have discussed the feasibility
of using the correlation function of Lyman-$\alpha$ clouds
seen along the lines-of-sight of neighboring high-redshift quasars
to distinguish between a low-density model and a flat model
with dark energy.  Fig.~\ref{fig:AP} shows the
sensitivity of this technique to $w$; whether or not it
has the power to probe the nature of the dark energy
remains to be seen.

The comoving volume element is the basis of number-count
tests (e.g., counts of lensed quasars, galaxies,
or clusters of galaxies).  It is given in terms
of $r(z)$ and $H(z)$
\begin{equation}
f(z) \equiv {dV\over dz d\Omega} = r^2(z)/H(z)\,.
\end{equation}
Note too that
\begin{equation}
f(z) = {dF(z)\over dz}\qquad F(z) = \int_0^z\,
f(z)dz = {r(z)^3\over 3}\,.
\end{equation}

\begin{figure}
\includegraphics[height=2.5in, width= 3in]{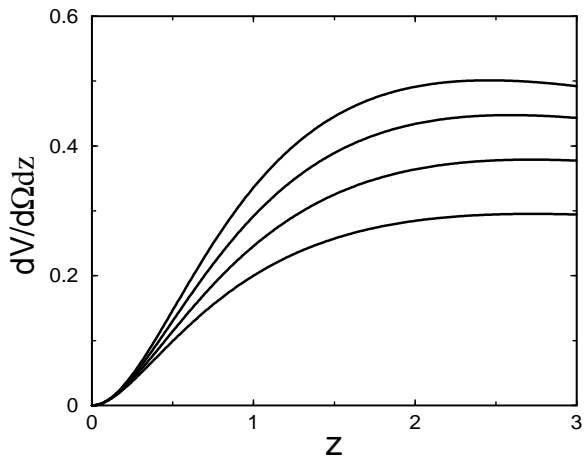}
\caption{Comoving volume element $f(z) = dV/d\Omega dz$
vs.\ redshift for constant $w=-1, -0.8, -0.6, -0.4$ (from top to
bottom).}
\label{fig:davis}
\end{figure}

The ability of these cosmological observables to probe the
dark-energy equation-of-state depends upon their
sensitivity to $w$.  To begin, consider the case of constant
$w$.  The sensitivity of $r(z)$, $H(z)$ and $f(z)$ to $w$
is quantified by

\begin{eqnarray}
{dr(z)\over dw} & = & -{3\over 2} \int_0^z\,
	{\Omega_X  H_0^2 (1+x)^{3(1+w)}
        \ln (1+x)\,dx \over H^3(x)} \\[0.2cm]
{df(z) \over dw} & = & {2r(z)\over H(z)}{dr\over dw} -
	{r(z)^2\over H(z)^2} {dH\over dw} \\[0.2cm]
{dH(z) \over dw} & = & {3\over 2} {\Omega_X H_0^2 (1+z)^{3(1+w)}
        \ln (1+z) \over H(z)}
\end{eqnarray}

\begin{figure}
\includegraphics[height=2.5in, width= 3in]{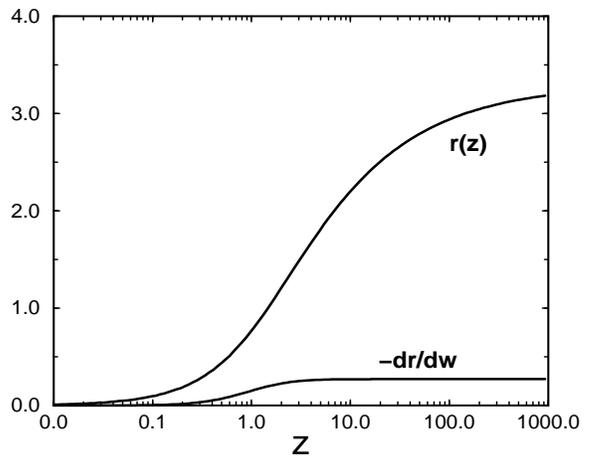}
\caption{$r(z)$ and $dr/dw$ as a function of $z$ (in units
of $H_0^{-1}$) for $\Omega_M = 0.3$ and $w=-1$.}
\label{fig:drdw}
\end{figure}

The comoving distance to an object at redshift $z$ and
its sensitivity to $w$ is shown in Fig.~\ref{fig:drdw}.
At small redshifts $r(z)$ is insensitive to $w$ for
the simple reason that {\em all} cosmological models
reduce to the Hubble law ($r = H_0^{-1} z$) for $z\ll 1$,

\begin{equation}
r(z) \rightarrow H_0^{-1}\left[ z -{3\over 4}z^2 -{3\over 4}
\Omega_X wz^2 +\cdots \right] \ \ {\rm for}\ z\ll 1
\label{eq:hubblelaw}
\end{equation}

At redshift greater than about five, the sensitivity of $r(z)$ to
a change in $w$ levels off because dark energy becomes an
increasingly smaller fraction of the total energy density,
$\rho_X /\rho_M \propto (1+z)^{3w}$.  As we shall discuss later,
the fact that $dr/dw$ increases monotonically with redshift means
that for measurements of fixed error, one would want to make the
measurement at the highest redshift possible in order to minimize
the uncertainty in the inferred value of $w$.

\begin{figure}
\includegraphics[height=2.5in, width= 3in]{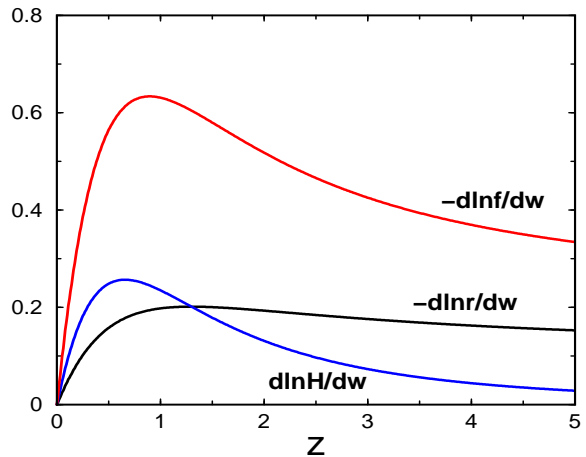}
\caption{The relative sensitivity of $r(z)$, $f(z)$, and $H(z)$
to a change in the constant value of $w$.}
\label{fig:sens2}
\end{figure}

Fig.~\ref{fig:sens2} shows the relative change in $r(z)$ and in
the comoving volume element $f(z)$ due to a change in $w$ as a
function of redshift.  The sensitivities of $r(z)$ and $f(z)$
peak at redshifts around unity.  The reason for decreased
sensitivity at small and large redshifts is as discussed just
above.

\begin{figure}[!ht]
\includegraphics[height=5.5cm, width=8cm]{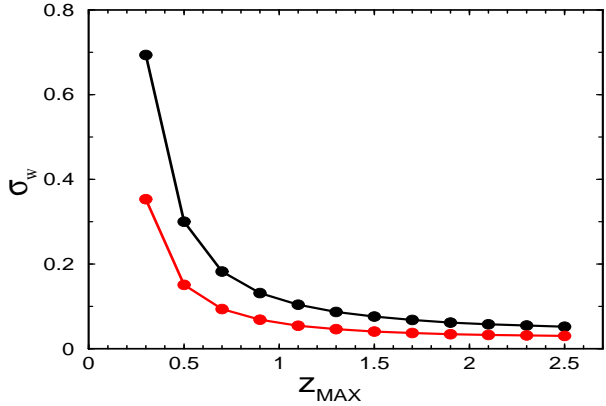}
\caption{$1\sigma$ accuracy in the determination of constant $w$ as a
function of maximum redshift probed $z_{\rm max}$ for a flat
Universe, 2000 SNe and marginalizing over the other parameter,
$\Omega_M$. The upper curve shows the uncertainties using the
fiducial SNAP dataset, while the lower curve shows uncertainties
obtained the mathematically optimal strategy (see
Sec.~\ref{optimal-sec}).  For all $z_{\rm max}$, 2000 SNe
were used, with a redshift distribution as shown in
Fig.~\ref{SNAP_hist.fig} for the upper curve.}
\label{w_vs_zmax.fig}
\end{figure}

As noted earlier, observations at redshifts $0\la z\la 2$
will be most useful in probing the dark energy. This fact is made
more quantitative in Fig.~\ref{w_vs_zmax.fig}, which shows the
accuracy of the determination of the equation-of-state $w$
(assumed constant) using 2000 SNe Ia, as a function of maximum
redshift probed $z_{\rm max}$.  For $0.2\la z_{\rm max}\la 1$,
the $1\sigma$ uncertainty $\sigma_w$ decreases sharply and then
levels, with little decrease for $z_{\rm max} \ga 1.5$.

\subsection{CMB anisotropy}

\begin{figure}
\includegraphics[height=2.5in, width= 3in]{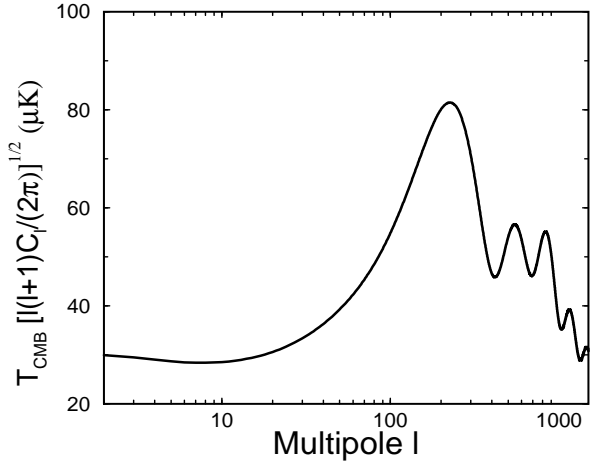}
\caption{COBE-normalized angular power spectrum of CMB anisotropy
for a flat model with $\Omega_Bh^2 = 0.02$, $\Omega_X=0.7$,
$h=0.65$ and $w=-1$, obtained using CMBFAST~\cite{CMBFAST}. The
acoustic peaks correspond to modes that at the moment of last
scattering are at maximum compression (odd) or rarefaction
(even).  }
\label{fig:CMB}
\end{figure}

The gravity-driven acoustic oscillations of the baryon-photon
fluid at the time of last scattering gives rises to a series of
acoustic peaks in the angular power spectrum of CMB anisotropy
(see Fig.~\ref{fig:CMB})~\cite{hu_sugiyama}.  The CMB is a
snapshot of the Universe at $z=z_{\rm LS}\simeq 1100$ and
the peaks correspond to different Fourier modes caught at maximum
compression or rarefaction, when the fluctuation in the
photon temperature is at an extremum.  The condition for this is
$k\eta_{\rm SH} \simeq n\pi$, where the odd (even) $n$ modes are
compression (rarefaction) maxima and $\eta_{\rm SH}$ is the sound
horizon:
\begin{eqnarray}
\eta_{\rm SH} & = & \int_0^{t_{\rm LS}} \, {v_s\,dt \over R(t)}
= \int_{z_{\rm LS}}^\infty {v_s (z^\prime )dz^\prime
	\over H(z^\prime )}\\
v_s^2  &=&  1/3\over 1+ {3\rho_B /4\rho_\gamma}
\end{eqnarray}

Modes captured at maximum compression or rarefaction
provide standard rulers on the last-scattering surface
with physical sizes $d \sim
\pi/k\,(1+z_{\rm LS}) \sim \eta_{\rm SH}/n\,(1+z_{\rm LS})$
Their angular sizes on the CMB sky are given by

\begin{eqnarray}
\theta_n & \sim & {\eta_{\rm SH}/n \over (1+z_{\rm LS})
        d_A(z_{\rm LS})}\\
d_A({\rm LS}) & = & (1+z_{\rm LS})^{-1}
	\int_0^{z_{\rm LS}} {dz^\prime \over H(z^\prime )}
\end{eqnarray}

This can be made more precise for the angular power spectrum.
The angular power at multipole $l$
is dominated by modes around $k\simeq l/\eta_{\rm LS}$,
and so the positions
of the peaks are given approximately by (see, e.g.,
Ref.~\cite{weinberg_peaks})
\begin{equation}
l_n = n\pi{\eta_{\rm LS} \over \eta_{\rm SH}}\,.
\end{equation}
For a flat Universe $\eta_{\rm LS}$ is just
the coordinate distance to the last-scattering surface
$r(z_{\rm LS})$.

\begin{figure}
\includegraphics[height=3in, width= 2.5in, angle=-90]{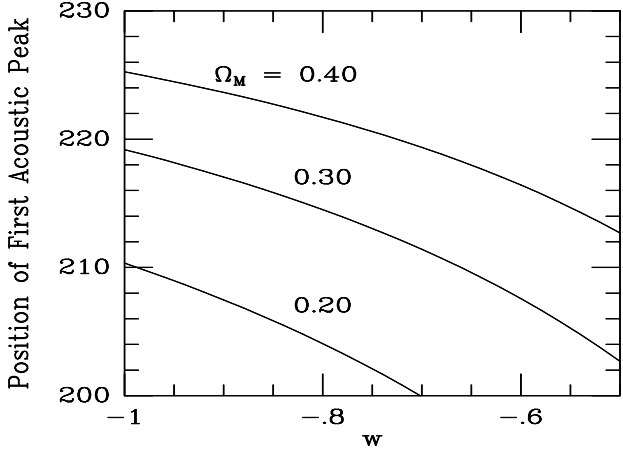}
\caption{The position of the first acoustic peak
as a function of $w$ for $\Omega_Bh^2=0.02$
and $\Omega_Mh^2 = 0.13$.}
\label{fig:peak}
\end{figure}

The positions of the acoustic peaks are the primary sensitivity
the CMB has to the equation-of-state of the dark energy
(see Fig.~\ref{fig:peak}).  Most of that sensitivity arises from
the dependence of the distance to the last-scattering
surface upon the time history of $w$.
Using the approximation above,
and taking into account the other important
cosmological parameters, it follows that

\begin{eqnarray}
{\Delta l_1\over l_1} &=& -0.084 \Delta w -0.23{\Delta \Omega_Mh^2
	\over \Omega_Mh^2}
        +0.09{\Delta \Omega_Bh^2\over \Omega_Bh^2}  \nonumber \\
       && \hspace{0.7cm} +0.089{\Delta \Omega_M \over \Omega_M}
        -1.25{\Delta \Omega_0 \over \Omega_0}
\label{eq:cmb_firstpeak}
\end{eqnarray}

\noindent around $w=-1$, $h=0.65$, $\Omega_M=0.3$, $\Omega_Bh^2 =0.02$
and $\Omega_0=1$.  Other features of the CMB power spectrum
(e.g., heights of the acoustic peaks and damping tail) can precisely
determine the matter density ($\Omega_Mh^2$) and the baryon
density ($\Omega_Bh^2$); therefore, for a flat Universe the main dependence
of the position of the acoustic peaks is upon $\Omega_M$ and $w$.
For $\Omega_M\sim 0.3$, $l_1$ is about three times more sensitive to
$\Omega_M$ than $w$.  Interestingly enough, the recent data from
the BOOMERanG and MAXIMA-1 experiments indicate that the first
peak is located at around $l\simeq 200$ \cite{Jaffe}, which
indicates a larger value of $w$, $w\sim -0.6$, than the supernova
data and suggests the dark energy may be something other than
a cosmological constant.  However, there is little statistical
significance to this result.

(The CMB angular power spectrum has additional sensitivity to
dark energy which is not captured by Eq.\ (\ref{eq:cmb_firstpeak}).  It
arises through the late-time ISW effect and the slight clumping
of dark energy, and mainly affects the low-order multipoles.
Because of the large cosmic variance at large scales, this
dependence is not likely to significantly enhance the ability of
CMB anisotropy to probe $w$.)

\subsection{Time-varying $w$}

There is no compelling reason to believe that the dark energy is
characterized by a constant $w$.  In particular, if
the dark energy is associated with an evolving scalar field then
the effective equation-of-state,
\begin{equation}
w(t) \equiv p_\phi /\rho_\phi = {{1\over 2}\dot\phi^2 - V(\phi ) \over
	{1\over 2}\dot\phi^2 + V(\phi )}\label{w_scalar.eq},
\end{equation}
is in general time-varying.
Thus, sensitivity to the value of $w(z)$ at a given $z$ is an
important measure of the probative power of a given test.  Needless
to say, in order to probe the variation of $w$ with redshift, one
has to perform measurements at different redshifts.  Thus,
CMB anisotropy and the age of the Universe cannot probe
this aspect of the dark energy; rather, they can
only measure some average value of $w$.

We now consider the effect of a change in $w$ at
redshift $z_*$; specifically, a change in $w$ over a
small redshift interval around $z=z_*$,
such that $\int \delta w(z) d\ln z
=1$.  The effect on $H(z)$ for $z>z_*$, which we denote
by the functional derivative $\delta H /\delta w(z)$, is

\begin{equation}
{\delta H(z) \over \delta w(z)} = {3\over 2}{z_* \over 1+z_*}
		{\Omega_X H_0^2 \exp \left [3\int_0^z (1+w)d\ln (1+z)
		\right ] \over H(z)}
\end{equation}

\noindent and zero for $z<z_*$. Note that the effect of
$\delta w(z)$ on the expansion rate is essentially to change it
by a fixed amount for $z>z_*$.

The effect on $r(z)$ and $f(z)=r(z)^2/H(z)$ follows by simple calculus:

\begin{eqnarray}
{\delta \ln r \over \delta w(z)} & = & {1\over r(z)}
	\int_0^z \left (-{\delta H\over \delta w}\right ){dz\over H(z)^2}\\
{\delta \ln f \over \delta w} & = & {2}{\delta \ln r\over
\delta w} - {1\over H(z)}{\delta H \over \delta w}
\end{eqnarray}

\begin{figure}[!ht]
\includegraphics[height=2.5in, width=3in]{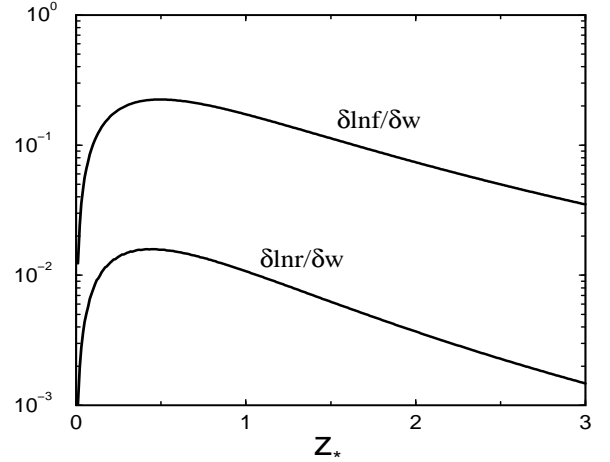}
\caption{The relative sensitivity of the comoving
distance $r(z)$ and the comoving volume element $f(z)$
to a localized change in the value of $w$ at redshift $z_*$
characterized by $\int \delta w(z) d\ln z = 1$.}
\label{fig:sens1}
\end{figure}

The sensitivity of $r(z)$ and $f(z)$ to a localized change in $w$
is shown in Fig.~\ref{fig:sens1}, where we take $z_* = 0.9z$.
Both $r(z)$ and the comoving volume element are insensitive to
the value of $w(z)$ at small redshift (since $r$ and $H$
are insensitive to the form of the dark energy) and at large
redshifts (because $\rho_X/\rho_M$ decreases rapidly).
They are most sensitive to $w(z)$ over the redshift range $z\sim
0.2 - 1.5$, with the sweet spot being at $z\approx 0.4$.

As discussed in Ref.~\cite{hutererturner}, measurements of $r(z)$
can in principle be used to reconstruct the equation-of-state (or
scalar-field potential in the case of quintessence).  The
reconstruction equation for $w(z)$ is
\begin{equation}
1+w(z) = {1+z\over 3}\, {3H_0^2\Omega_M(1+z)^2 + 2(d^2r/dz^2)/(dr/dz)^3
	\over H_0^2\Omega_M(1+z)^3-(dr/dz)^{-2}}.
\label{eq:reconw}
\end{equation}

This equation can be used to illustrate yet again the difficulty
of probing the dark energy at high redshift.  Suppose that $r(z)$
and its derivatives are measured very accurately and that the
only uncertainty in reconstructing $w(z)$ is due to $\Omega_M$.  The
uncertainty in $w(z)$ due to the uncertainty in $\Omega_M$ can be
obtained by differentiating Eq.\ (\ref{eq:reconw}) with respect
to $\Omega_M$:

\begin{eqnarray}
\Delta w(z) & = & {-(1+z)^3 \over \Omega_X \exp \left[
        3\int (1+w)\,d\ln (1+z) \right]}\,\Delta \Omega_M \\
        & \rightarrow & {-(1+z)^{-3w}\over \Omega_X}\,\Delta
        \Omega_M\ \ \ ({\rm const\ }w)
\end{eqnarray}

Therefore, the uncertainty in $w(z)$ increases with redshift
sharply, as $(1+z)^{-3w}$. This happens because $w<0$ and the
dark energy constitutes an increasingly smaller fraction of the
total energy at high redshift.

The reconstruction equations based upon number counts can simply
be obtained by substituting $[3F(z)]^{1/3}$ for $r(z)$ in Eq.\
(\ref{eq:reconw}).  Since the expansion history $H(z)$ can in
principle be obtained from measurements of $f(z)$ and $r(z)$
(number counts and Hubble diagram), or from $r(z)$ and $r(z)H(z)$
(Hubble diagram and Alcock-Paczynski test), with a sense of
great optimism we write the reconstruction equation based upon a
determination of $H(z)$:

\begin{equation}
1+w(z) = {1\over 3} {2(1+z)H^\prime(z) H(z) -3H_0^2(1+z)^3\Omega_M
        \over H^2(z) - H_0^2\Omega_M(1+z)^3}
\end{equation}

\noindent which follows from

\begin{eqnarray}
dr(z) & = & -dt/a(t) = dz/H(z).
\end{eqnarray}

\noindent This reconstruction equation has the virtue of
depending only upon the first derivative of the empirically
determined quantity.

\subsection{Summing up}

In sum, the properties of the dark energy are best revealed by probes
of the low-redshift ($z\sim 0.2 - 2$) Universe -- SNe Ia, number-counts and
possibly the Alcock-Paczynski test. The CMB has an
important but more limited role to play since it can only
probe an average value of $w$. SNe Ia are currently the
most mature probe of the dark energy, and already impose
significant constraints on $w$ \cite{garnavich, perl_tur_white,
perlmutter-1999}, $w< -0.6$ (95\% CL).  The efficacy of any
of these tests will depend critically upon the identification
and control of systematic error (more late).

The classical cosmological tests that involve $r(z)$ alone have
the virtue of only depending upon $\Omega_M$, $\Omega_X$ and $w$,
which can be reduced to two parameters ($\Omega_M$ and $w$) with
a precision measurement of $\Omega_0$ from the CMB.
CMB anisotropy on the other hand depends upon a large number
of parameters (e.g., $\Omega_Bh^2$, $h$, $n$, $dn/d\ln k$,
ionization history, etc).  The number-count tests can
also depend upon the growth of structure which brings in
other parameters that affect the shape of the power spectrum
(e.g. $n$, $\Omega_Bh^2$, $h$).

In the remainder of this paper we pay special
attention to SNe Ia, and in particular consider how well the dark
energy could be probed by a high-quality dataset provided by
the proposed satellite mission SNAP~\cite{SNAP}. As the
fiducial dataset, we consider a total of 2000 SNe
Ia with individual statistical
uncertainties of $0.15$ mag (the impact of systematic
uncertainties to this dataset was studied in Refs.~\cite{SNAP,
weller, astier}). The bulk of the SNe are assumed to have
$0.2<z<1.2$, with about a hundred at $1.2<z<1.7$
and another two hundred or so at $z<0.2$.  The
low-z sample is expected from near-future ground searches, such
as the Nearby Supernova Factory~\cite{SN_factory}.

\begin{figure}[!ht]
\includegraphics[height=5.5cm, width=8cm]{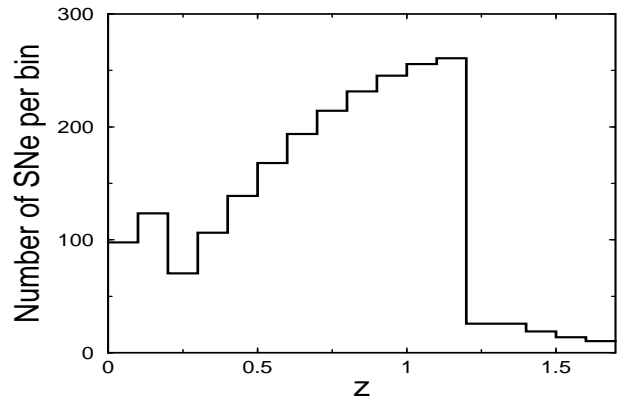}
\caption{Histogram of projected SNe Ia distribution from SNAP.
The number of SNe at $z>1.2$ is smaller because spectra of SNe at
such high $z$ are redshifted into the infrared region, where
observations are more difficult. About 200 SNe at $z<0.2$ are
assumed to be provided by ground-based SNe searches.
\label{SNAP_hist.fig}}
\end{figure}

The number-count technique can be implemented in a variety of
ways --- for example, halos of a fixed mass~\cite{davis},
clusters of galaxies of fixed mass~\cite{holder}, and
gravitationally lensed quasars~\cite{coorayhuterer}.  All of
these methods, however, are susceptible to redshift evolution of
the objects in question, as well as considerable uncertainties in
theoretical modeling.

Unless otherwise indicated, we use the Fisher-matrix formalism
throughout to estimate
uncertainties~\cite{Fisher-Jungman,Fisher-Tegmark}. In several
instances we have checked that the values obtained
agree well with those using Monte-Carlo simulation.
The fiducial cosmological model is $\Omega_M=1-\Omega_X=0.3$,
$w=-1$, unless otherwise indicated.

\section{Constraints on (constant) $w$}\label{w-sec}

To begin, we assume that the equation-of-state of the dark energy
does not change in time, $w(z)=w={\rm const}$. Not only does this
hold for models with truly constant $w$ (vacuum energy, domain
walls and cosmic strings, etc.) but models with time-variable
equation-of-state can have $w\approx {\rm const}$ out to $z\sim
2$.

\subsection{SNe Ia and CMB}\label{SNeCMB-subsec}

\begin{figure}[!ht]
\includegraphics[height=5.5cm, width=8cm]{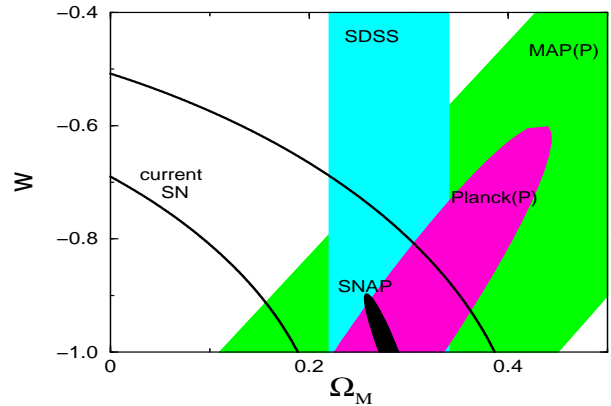}
\caption{Projected SNAP constraint compared to
those projected for MAP and Planck (with
polarization information) and SDSS (MAP, Planck and SDSS
constraints are from Ref.~\cite{hu_obs_99}). Also shown are the
present constraints using a total of 54 SNe Ia. All constraints
assume a flat Universe and $\Omega_M=1-\Omega_X=0.28$, $w=-1$ as
fiducial values of the parameters. All contours are 68\% cl,
and were obtained using the Fisher-matrix analysis.
\label{mat_w.SN_CMB.fig}}
\end{figure}

Fig.~\ref{mat_w.SN_CMB.fig} shows that a supernova program, such
as SNAP~\cite{SNAP}, will enable very accurate measurement of
$w$: $\sigma_w\approx 0.05$, after marginalization over
$\Omega_M$ and assuming a flat Universe.  This figure also shows
constraints anticipated from the Sloan Digital Sky Survey (SDSS)
and MAP and Planck satellites (with polarization information). As
expected, the fact that the dark energy is smooth on observable
scales implies that its properties cannot be probed well by
galaxy surveys. 

The CMB, on the other hand, is weakly sensitive to the dark
energy, mainly through the dependence of the distance to the
surface of last scattering upon $w$.  The orientation of the CMB
ellipses is roughly predicted from Eq.\ (\ref{eq:cmb_firstpeak}),
indicating that this equation captures most of the CMB dependence
upon dark energy.

The CMB provides only a
{\em single} measurement of the angular-diameter distance to the
surface of last scattering, albeit an accurate one. In the case
of Planck, the angular-diameter distance to the last-scattering
surface is measured to
0.7\%~\cite{eisen-private}. Fig.~\ref{mat_w.SN_CMB.fig}
illustrates that ultimately CMB is not likely to be as precise as
a well-calibrated SNe dataset. It does provide complementary
information and a consistency check. However, combining SNAP and
Planck improves the SNAP constraint only by about 5-10\%.

The uncertainty in the determination of $w$ varies as a
function of the central value of this parameter. As $w$ increases
from $-1$, the SNe constraint becomes weaker: the variation of
dark energy with redshift becomes similar to that of matter, and
it is more difficult to disentangle the two components by
measuring the luminosity distance.  For example, for $w=-0.7$ and
keeping $\Omega_M=0.3$, the constraints on these two parameters
from SNAP deteriorate by 10\% and 50\% respectively relative to
the $w=-1$ case.  On the other hand, the CMB constraint becomes
somewhat {\it stronger} with increasing $w$ because the ISW
effect increases (see Fig.\ 5 in Ref.~\cite{hu_obs_99}).

\subsection{Number-counts}\label{numcount-subsec}

\begin{figure}[!ht]
\includegraphics[height=5.5cm, width=8cm]{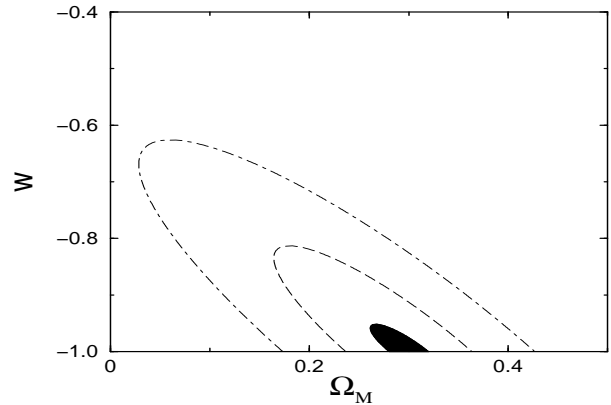}
\caption{Constraints in the $\Omega_M$-$w$ plane using
galaxy-halo counts from the DEEP survey~\cite{davis}. 
Inner most region shows the constraint assuming
Poisson errors only, while the
outer two regions assume an additional, irreducible
uncertainty of 10\% and 20\% (per bin) in the comoving
number density of halos due to evolution.
All regions are 68\% cl.
\label{mat_w.halos.fig}}
\end{figure}

Davis and Newman~\cite{davis} have argued that the comoving
abundance of halos of a fixed rotational speed (nearly fixed
mass) varies weakly with the cosmological model and can be
calibrated with numerical simulations, leaving mostly the
dependence on the volume element~\cite{davis}. We follow these
authors in assuming 10000 galaxy halos divided into 8 redshift
bins at $0.7<z<1.5$.  The redshift range for the DEEP survey
roughly corresponds to the redshift range of greatest sensitivity
to dark energy.

Fig.~\ref{mat_w.halos.fig} shows the constraints obtained using
the Fisher-matrix formalism assuming Poisson errors only, and
then allowing for an additional 10\% or 20\% error per bin for
the uncertainty in the evolution of the comoving halo density.
Assuming no uncertainty in the comoving halo density, the error
ellipse is similar to that of SNAP. However, allowing for a
modest uncertainty due to evolution (10 or 20\%), the size of the
error ellipse increases significantly. Finally, we note that any
probe sensitive primarily to $dV/d\Omega dz$ will have its error
ellipse oriented in the direction shown.

\begin{figure}[!th]
\includegraphics[height=10cm, width=8cm]{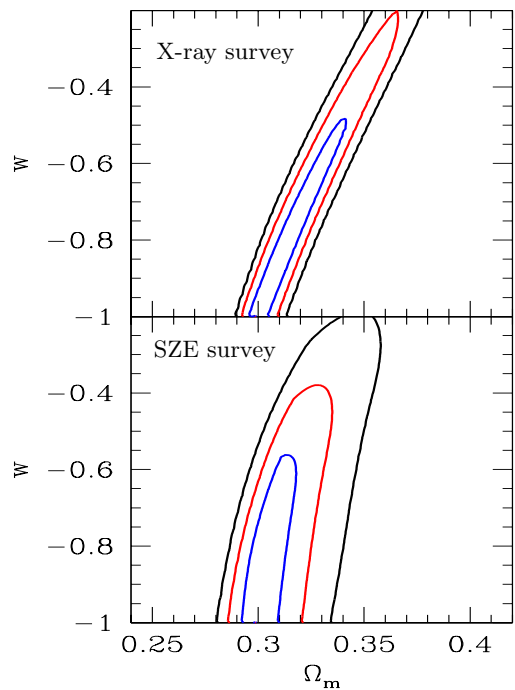}
       \put(-155, 250){X-ray survey}
       \put(-156, 137){SZE survey}
\caption{Projected one, two and three $\sigma$
constraints on $\Omega_M$ and $w$ in a flat Universe
using counts of galaxy clusters (adopted from Ref.~\cite{holder})
for an x-ray selected sample of thousand clusters (top panel)
and a Sunyaev-Zel'dovich selected sample of hundred
clusters (bottom).
\label{clusters.fig}}
\end{figure}

While clusters are simpler objects than galaxies, they are ``rare
objects'' and their abundance depends exponentially upon the
growth of density perturbations and varies many
orders-of-magnitude over the redshift range of
interest~\cite{bahcall}.  The sensitivity to the growth factor
outweighs that of the cosmological volume, and the error ellipses
for the cluster number-count test are almost orthogonal to the
halo-number count test (see Fig.~\ref{clusters.fig}).  The
information provided is thus complementary to halo counts and SNe
data.

Because of the exponential dependence of the abundance, control
of the systematic and modeling errors is critical.  Especially
important is accurate determination of cluster masses (use of
weak-gravitational lensing to determine cluster masses might be
very useful~\cite{gilholder}).  Shown in
Fig.~\ref{clusters.fig} are the estimated constraints for
sample of a hundred clusters with $0 < z < 3$ selected in a
future Sunyaev-Zel'dovich survey and a thousand clusters with $0
< z < 1$ selected in a future x-ray survey~\cite{holder}.  These
cluster constraints are comparable and complementary to those of
the halo counts when a 20\% uncertainty in halo-number evolution
is taken into account.

\section{Probing the Time-history of Dark Energy}\label{wz-sec}

Although some of the models for the dark energy, such as the
vacuum energy, cosmic defects and some quintessence models
produce $w={\rm const}$ (at least out to redshifts of a few), the
time-variation of $w$ is a potentially important probe of the
nature of dark energy. For example, evolving scalar field models
generically time-variable $w$.  Moreover, in some cases
(e.g. with PNGB scalar field models~\cite{coble} and some tracker
quintessence models~\cite{zlatev}) $w(z)$ can exhibit significant
variation out to $z\sim 1$.

\subsection{Constraining the redshift dependence of $w$}

Given a dark-energy model it is easy to compute $w(t)$ and from
it the prediction for $r(z)$.  There is little theoretical guidance
as to the nature of the dark energy, so we seek ways to
parameterize $w(z)$ as generally as possible. A further
complication is the degeneracy of $w(z)$ with $\Omega_M$ and
$\Omega_X$. To make useful progress, we assume that by the time a
serious attempt is made to probe the rate of change of $w$,
$\Omega_M$ and $\Omega_X$ will be measured accurately (e.g.,
$\Omega_M+\Omega_X$ from CMB anisotropy, and $\Omega_M$ from
large-scale structure surveys).

\subsection{Case I: $w(z)=w_1 +w'(z-z_1)$}

\begin{figure}[th]
\includegraphics[height=5.5cm, width=8cm]{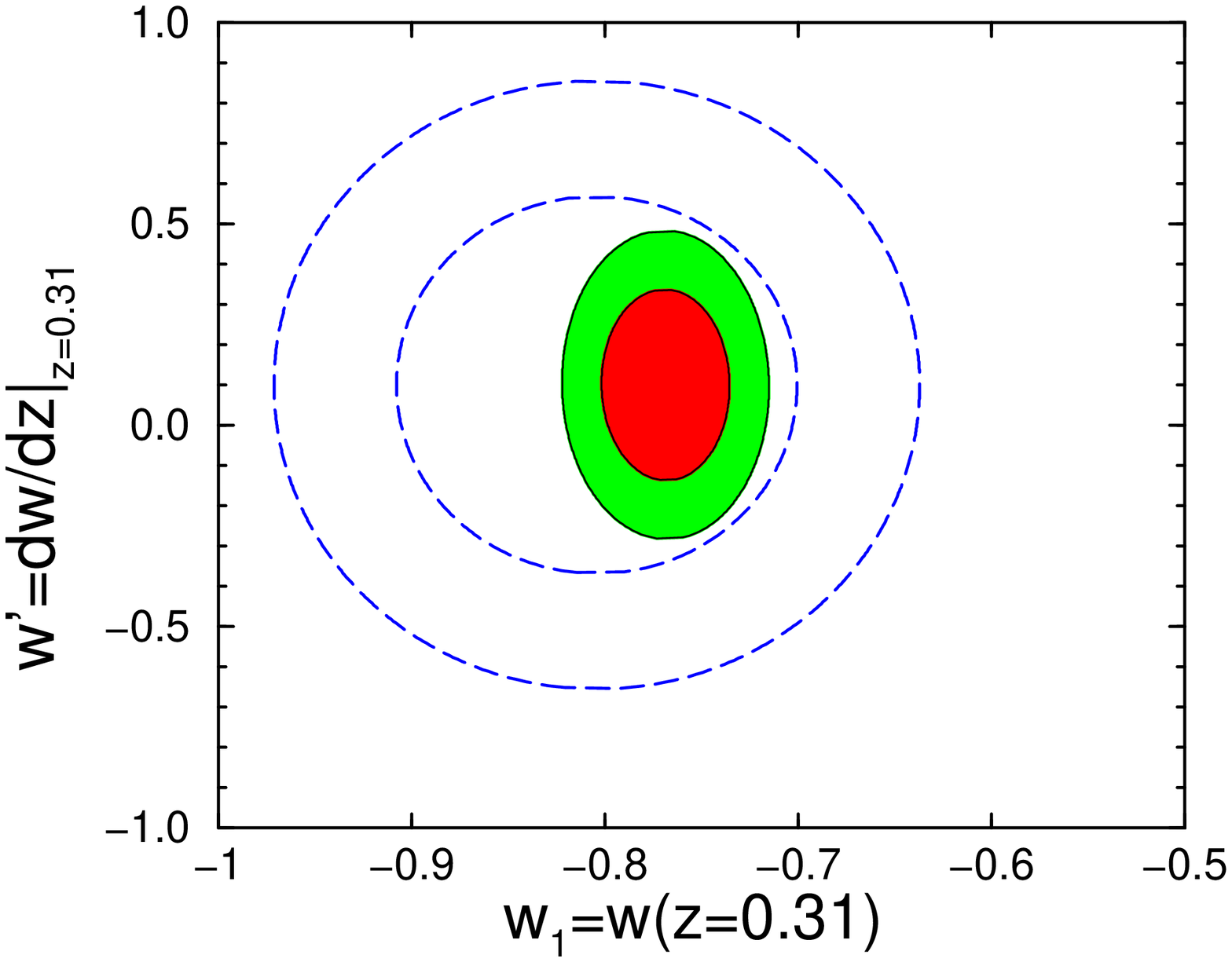}
\includegraphics[height=5.5cm, width=8cm]{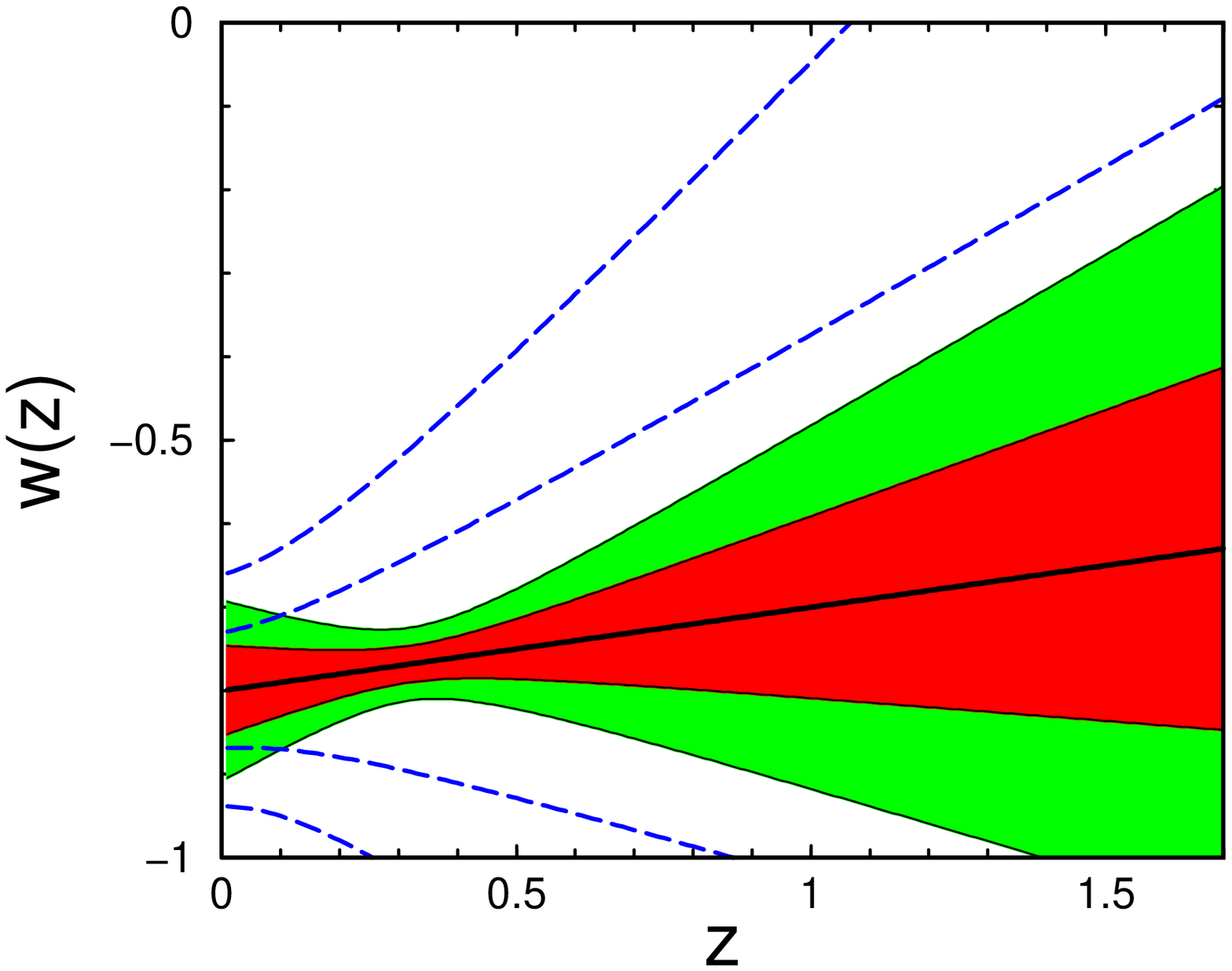}
\caption{Here $w(z)$ was Taylor-expanded around $z_1=0.31$ with
fiducial model $w(z)=-0.8+0.1\,z$. The top panel shows 68\% and 95\% cl
constraints in the $w_1$-$w'$ plane. The bottom panel shows the
same constraint in the $w$-$z$ plane, with the fiducial model
(heavy line) and confidence regions (shaded).  The broken lines
in both panels should the effect of assuming a Gaussian
uncertainty of $0.05$ in $\Omega_M$.
\label{w_wprim.fig}}
\end{figure}

The simplest way to parameterize the rate of change of $w$ is to
write the first-order Taylor expansion

\begin{equation}
w(z)=w_1 +w'(z-z_1),
\label{eq:w_linear}
\end{equation}

\noindent where $w_1=w(z=z_1)$ and $w'\equiv (dw/dz)_{z=z_1}$ are
constants and $z_1$ is the redshift around which we expand
(chosen according to convenience or theoretical
prejudice).  The energy density in the dark component is then
given by

\begin{equation}
\rho_X (z) = \rho_X(0) (1+z)^{3(1+w_1-w'(1+z_1))}\exp(3w'\,z).
\end{equation}

Using the Fisher-matrix formalism, we determine the error
ellipses in the $w_1$-$w'$ plane.  We choose $z_1$
so that $w_1$ and $w'$ become uncorrelated, (how to
do this analytically is shown in Ref.~\cite{Eisenstein}). For
uncorrelated $w_1$ and $w'$, the constraint to $w(z)$
follows by computing

\begin{equation}
\sigma_{w(z)}=
\left [
 \sigma_{w_1}^2 + \sigma_{w'}^2\,(z-z_1)^2
\right ]^{1/2}.
\end{equation}

Fig.~\ref{w_wprim.fig} illustrates the error ellipse for $w_1$ and $w'$
(top panel) and the constraint to $w(z)$ (bottom panel). As we
discussed in Sec.~\ref{prelim-sec}, cosmological observations
have diminishing leverage at both high and low redshift, which is
reflected in the narrow ``waist'' at $z\sim 0.35$, and this is the
sweet spot in sensitivity to $w(z)$ (see Fig.~\ref{fig:sens1}).

The uncertainty in the slope, $\sigma_{w'}=0.16$, is about 8
times as large as that in $w(z_1)$, $\sigma_{w_1}=0.02$. 
Despite the relatively large uncertainty in $w'$, this analysis
may be useful in constraining the dark-energy models.

Finally, we also show in Fig.~\ref{w_wprim.fig} the significant
effect of a Gaussian uncertainty of $0.05$ in $\Omega_M$; it roughly
doubles $\sigma_{w_1}$ and $\sigma_{w'}$ and moves the value
of $z_1$ that decorrelates the two parameters  to less than zero.

\subsection{Case II: $w(z)=w_1-\alpha \ln[(1+z)/(1+z_1)]$}

There are other ways to parameterize the variation of $w(z)$ with
redshift.  Efstathiou \cite{efstathiou} argues that many
quintessence models produce equation-of-state ratio that is well
approximated by $w(z)=w_0-\alpha \ln(1+z)$ with $w_0$ and
$\alpha$ constants. We generalize this by expanding
around an arbitrary redshift $z_1$

\begin{equation}
w(z)=w_1-\alpha \ln\left ({1+z\over 1+z_1}\right ).
\end{equation}

\noindent Here, the energy density in the dark energy evolves as

\begin{eqnarray}
\rho_X (z)&=&\rho_X(0) (1+z)^{3(1+w_1+\alpha \ln(1+z_1))}\times \\[0.1cm]
&&\exp\left [-{3\over 2}\,\alpha\ln^2(1+z)\right ].
\end{eqnarray}

As with the Taylor expansion, we have a 2-parameter form for
$w(z)$ and, using the supernova data, we examine the constraints
that can be imposed on $w_1$ and $\alpha$. We again choose $z_1$
so that $w_1$ and $\alpha$ are decorrelated; this occurs for
$z_1=0.30$.

\begin{figure}[!ht]
\includegraphics[height=2.5in, width=3in]{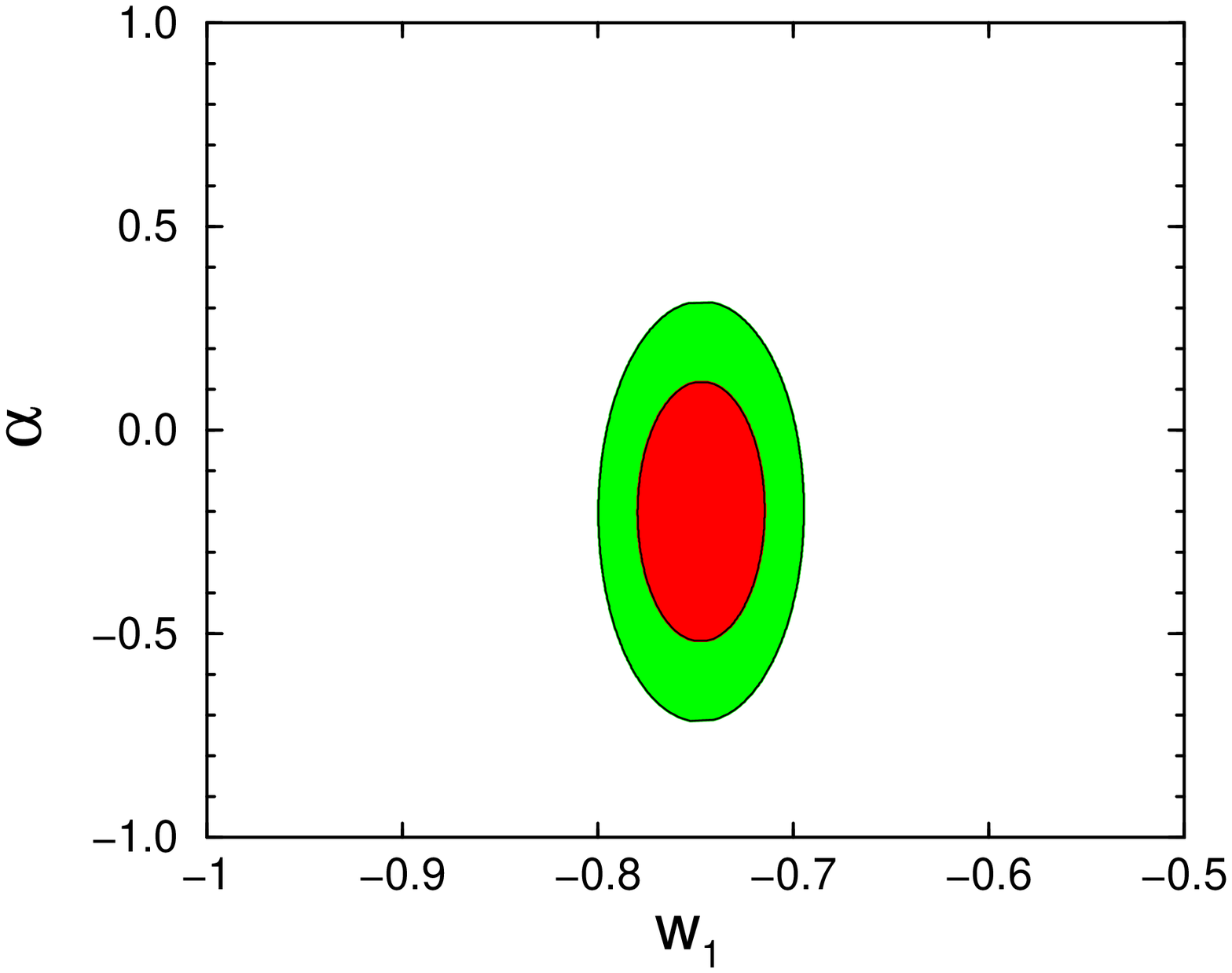}
\includegraphics[height=2.5in,  width=3in]{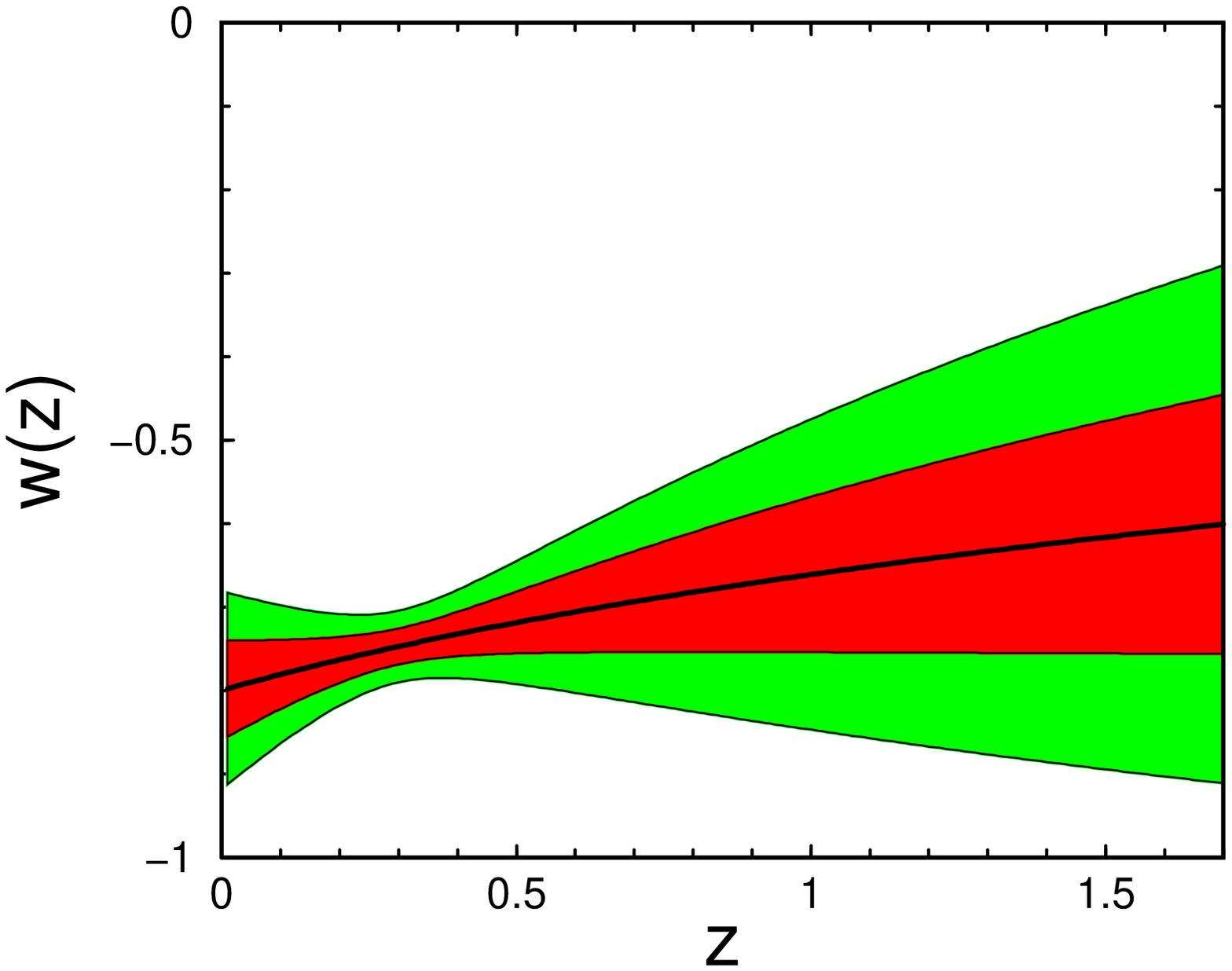}
\caption{Here the dark energy is parameterized by $w(z)=w_1-\alpha
\ln[(1+z)/(1+z_1)]$, with $w_1=-0.75$ and $\alpha=-0.2$. 
The top panel shows 68\% and 95\% cl constraints in
the $w_1$-$\alpha$ plane. The bottom panel shows the same
constraint in the $w$-$z$ plane, with the fiducial model (heavy
line) and 68\% and 95\% cl confidence region (shaded).
\label{w_alpha.fig}}
\end{figure}

Fig.~\ref{w_alpha.fig} shows 68\% and 95\% cl constraint regions
in the $w_1$-$\alpha$ plane (top panel) and $w$-$z$ plane
(bottom).  The fiducial model ($w_1=-0.75$, $\alpha=-0.2$) is
chosen to produce $w(z)$ similar to that from linear expansion
(Case I).  The uncertainty in parameter determination is
$\sigma_{w_1}=0.02$ and $\sigma_{\alpha}=0.21$.  The bottom panel
of this figure shows that using the logarithmic expansion we
obtain similar constraints to $w(z)$ as with the linear
expansion.  This is not surprising, as near the leverage point
$z_1\approx 0.3$, the two expansions are essentially equivalent
with $\alpha=(1+z_1)w'$ and $\sigma_{\alpha}=(1+z_1)\sigma_{w'}$,
which is consistent  with our results.

\subsection{Case III: Constant $w$ in redshift bins}

An even more general way to constrain $w(z)$ is to parameterize it by
constant values in several redshift bins, since no
particular form for $w(z)$ need be assumed.  Of course, more
redshift bins lead to weaker constraints in each bin.

We divide the SNAP redshift range into $B$ bins centered at
redshifts $z_i$ with corresponding widths $\Delta z_i$ and
equation-of-state ratios $w_i$ ($i=1, \ldots, B$). The energy
density of the dark component evolves as ($z_{j-1}<z<z_j$)

\begin{eqnarray}
\rho_X (z)&= &\rho_X(z=0) \,\prod_{i=1}^{j-1} \left (\frac{1+z_i+\Delta
z_i/2}{1+z_i-\Delta z_i/2} \right )^{3(1+w_i)}\times \nonumber \\[0.1cm]
&&\left (\frac{z}{1+z_{j}-\Delta z_{j}/2} \right )^{3(1+w_j)}.
\end{eqnarray}

\noindent To obtain the constraints using this approach, we again
employ the Fisher matrix formalism, treating $w_i$ as the parameters
to be determined.

\begin{figure}[!ht]
\includegraphics[height=2.5in,width=3in]{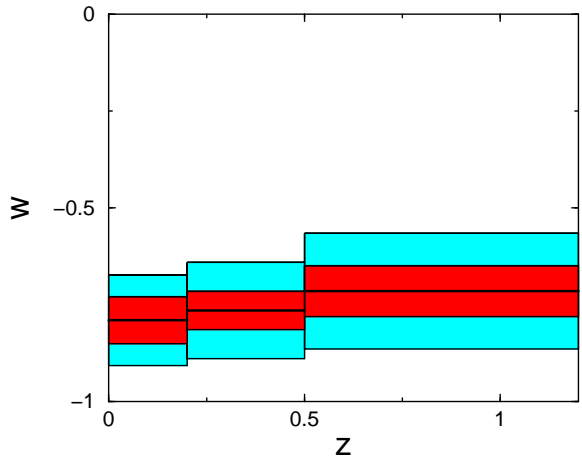}
\caption {Here $w(z)$ is parameterized by constant values in
redshift bins. The outer region shows 68\% cl constraints
corresponding to each redshift bin. The inner region shows 68\%
cl constraints when, in addition, a Gaussian prior is imposed
that penalizes models with a large change in $w$ between two
adjacent bins.}
\label{wz_param.fig}
\end{figure}

Fig.~\ref{wz_param.fig} shows constraints on $w(z)$ when $w$ is
parameterized by values in three redshift bins whose widths are
chosen so that the uncertainty in each is about the same. Precise
knowledge of $\Omega_M$ and $\Omega_X$ was assumed.

The constraints are not strong ($\sigma_w\approx 0.12$)  in part
because the values of $w$ in adjacent bins are uncorrelated.
Most realistic models with time-dependent equation-of-state have
$w(z)$ that varies slowly (or doesn't vary at all) out to $z\sim
1$. Therefore, we also show results when a Gaussian prior is
imposed that penalizes models with large change in $w$ between
two adjacent bins ($\sigma_w = 0.10$ for change in $\Delta w_i$
between adjacent bins).
The $1\sigma$ constraints improve by more than a factor of two.

\subsection{Non-parametric Reconstruction}

The most general approach is the direct reconstruction of $w(z)$
from the measured luminosity distance -- redshift relation
provided by the SNe Ia data~\cite{hutererturner, chiba, saini,
chiba2}.  This method is non-parametric and no assumptions about
the dark energy or its equation-of-state are needed.  This is
also the most challenging approach, since the reconstructed
potential and equation-of-state ratio will depend on first and
second derivatives of the distance with respect to redshift,
cf.\ Eq.~(\ref{eq:reconw}). This leads to a fundamental problem:
even very accurate and dense measurements of $r(z)$ allow great
freedom in $r'\equiv dr/dw$ and $r''\equiv d^2r/dw^2$, because
they themselves are not probed directly.

To address this problem, various authors have advocated
polynomials and Pad\'{e} approximants~\cite{hutererturner} and
various fitting functions~\cite{saini, chiba2, weller} to
represent $r(z)$ and thereby reduce the inherent freedom in $r'$
and $r''$.

\begin{figure}[!ht]
\includegraphics[height=5.5cm, width=8cm]{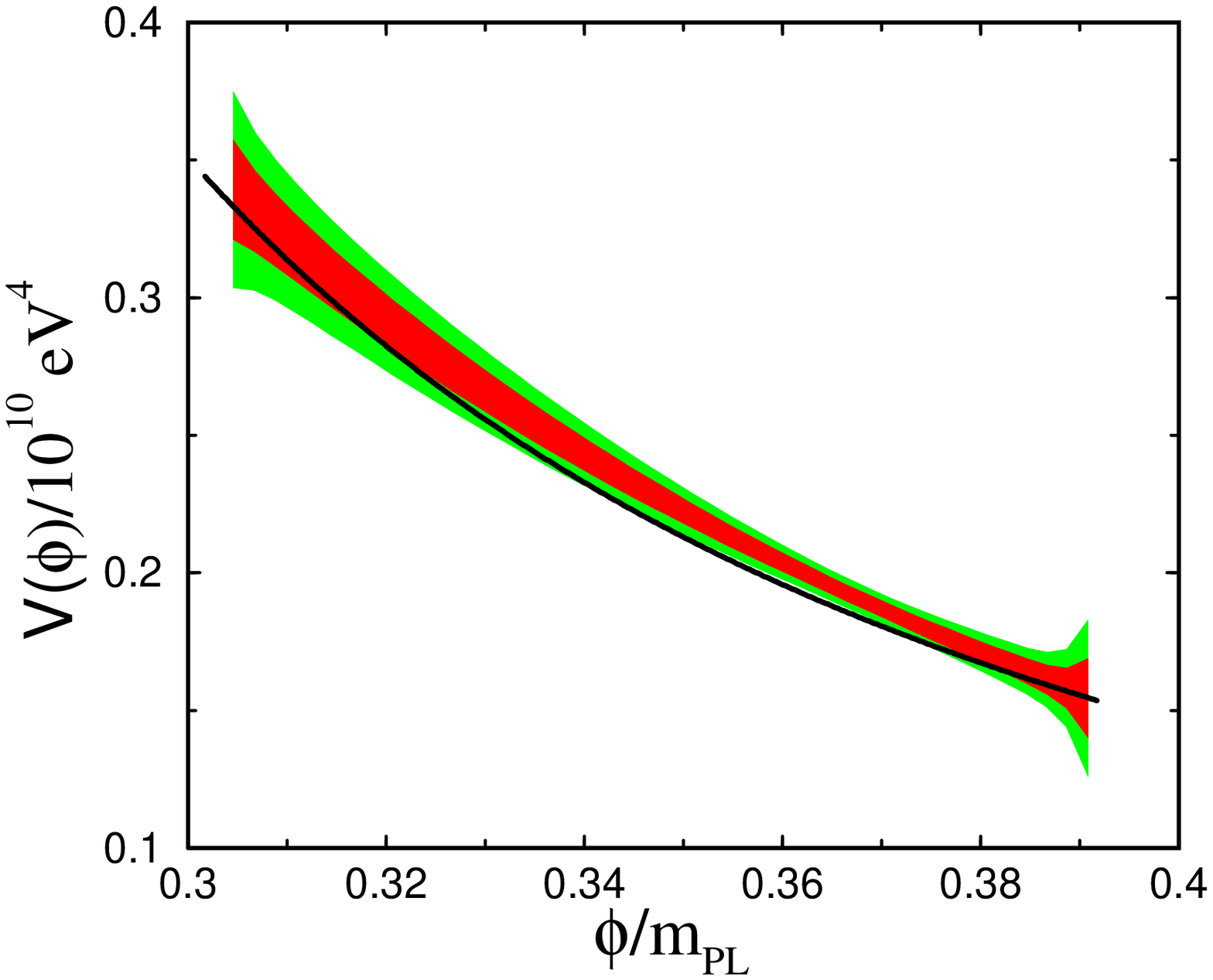}
\includegraphics[height=5.5cm, width=8cm]{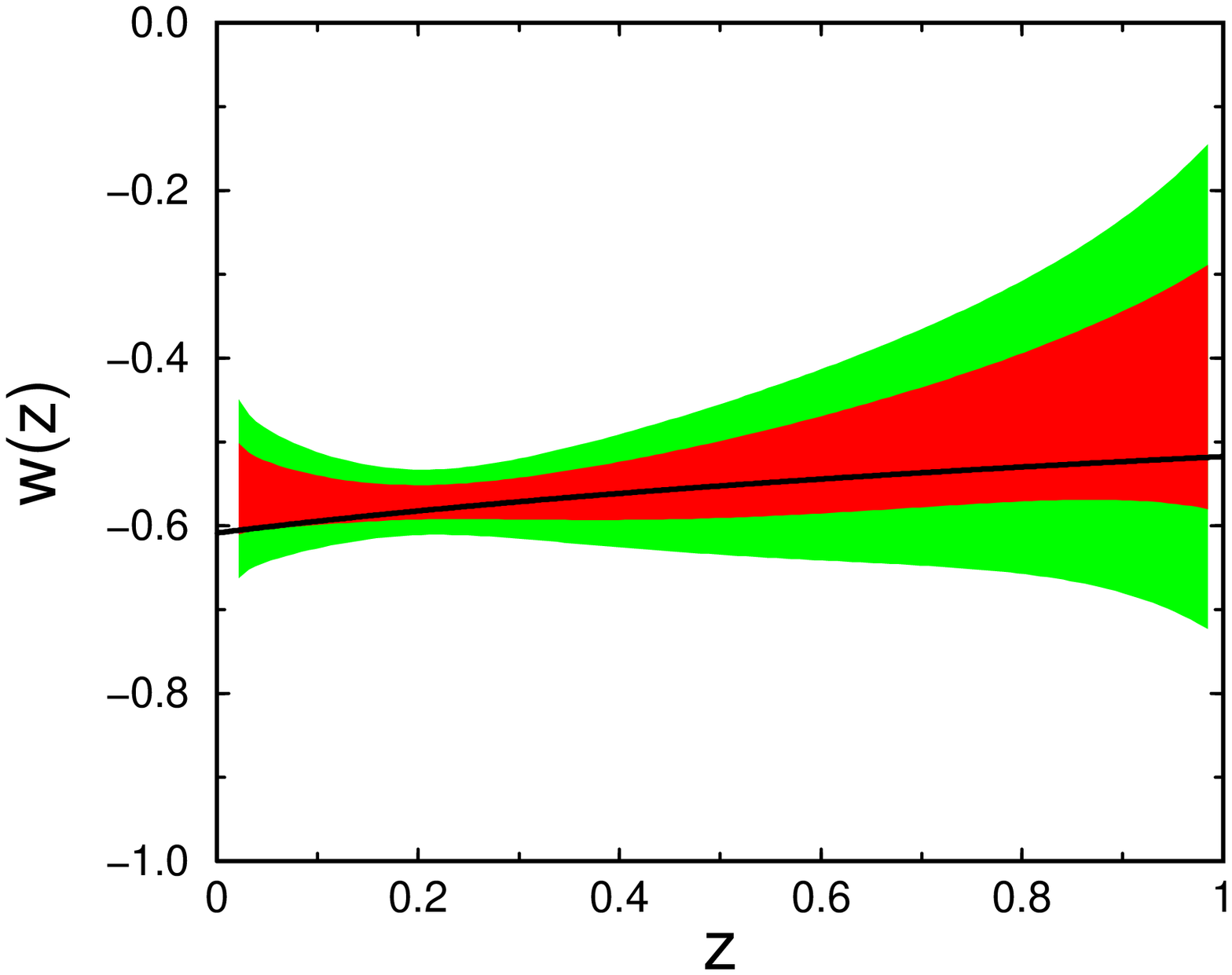}
\caption{Reconstruction of the quintessence model with potential
$V(\phi)=M^4 [\exp(\mpl/{\phi})-1]$~\cite{zlatev} and
$\Omega_X=0.50$.  The solid line is the input model, and the
shaded regions are the 68\% and 95\% confidence, produced
from Monte Carlo simulation of 2000 SNe
uniformly distributed out to $z=1.5$ with individual
uncertainties of $0.15$ mag (7\% in distance).  A
three-parameter Pad\'{e} approximant fit to $r(z)$
was used.  In the lower panel the reconstruction is shown
as $w(z) = ({1\over 2}\dot\phi^2 -V)/({1\over 2}\dot\phi^2 -V)$.
\label{reconstr.fig}}
\end{figure}

In Fig.~\ref{reconstr.fig}, we show the simulated
reconstruction of the quintessence model with potential
$V(\phi)=M^4[\exp(\mpl/{\phi})-1]$~\cite{zlatev} and
$\Omega_X=0.50$.  We assumed 2000 SNe uniformly distributed out
to $z=1.5$ with individual uncertainties of $0.15$ mag. Data were
fit by a three-parameter Pad\'{e} approximant of the form

\begin{equation}
r(z)=\frac{z(1+a\,z)}{1+b\,z+c\,z^2}.
\end{equation}

We have also tried other fitting functions that have been
suggested~\cite{saini, chiba2, weller}, as well as a piecewise
cubic spline with variable tension. We find that all are able to
fit the predicted form for $r(z)$ well (about 0.2\%
accuracy). However, a good fit is not the whole story -- $r'(z)$
and $r''(z)$ are equally important -- and the small bumps and
wiggles between the between the fit and the actual form predicted
by the dark-energy model are important because they lead to
reconstruction error.

In sum, non-parametric reconstruction is very challenging, and an
oxymoron: as a practical matter the data must be fit by a smooth
function. Nevertheless, in the absence of a handful of well
motivated dark-energy models, reconstruction offers a more
general means of getting at the time dependence of $w$ and the
very nature of the dark energy. Finally, it goes without saying
that the best way to test a  specific model is to use {\em it}
as a representation of the dark energy.

\subsection{Number Counts}

Probing $w(z)$ by number counts will involve all the difficulties
just discussed for SNe Ia, and the additional issue of separating
the evolution of the comoving density of objects
(galaxies or clusters) from the cosmological
effects of dark energy. To test the probative power of number
counts, we consider a cosmological probe that is primarily
sensitive to the volume element $dV/dzd\Omega$, such as the
galaxy-halo test using the DEEP survey~\cite{davis}.
In order to achieve comparable constraints to those provided by
SNe Ia, we find that $dV/dzd\Omega$ must be measured to 2-3\%
in each redshift bin.  Even with thousands of halos,
the accuracy in the number counts in each redshift bin
must be Poisson-limited -- a very challenging
goal when the ever-present uncertainties in theoretical
predictions of abundances of these objects are taken into
account.

\begin{figure}[!ht]
\includegraphics[height=5.5cm, width=8cm]{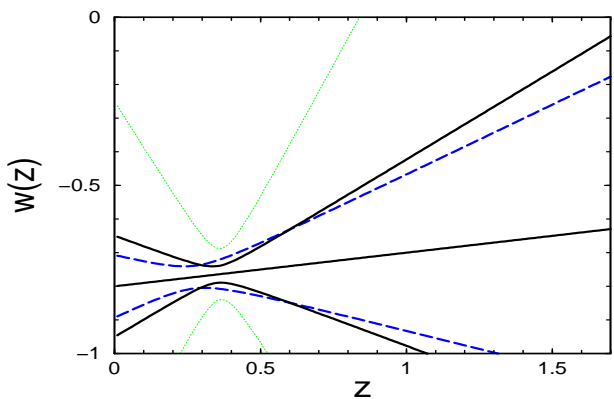}
\caption{The 95\% cl constraint on $w(z)$ when the dark energy is
parameterized by $w_1$ and $w'$ and the halo counts are divided
into 8 redshift bins with $0.7<z<1.5$ (solid lines) and 15
redshift bins with $0<z<1.5$ (dashed lines). The light dotted
lines show the result with 8 bins and $0.7<z<1.5$, but now with a
10\% (per bin) additional uncertainty due to the evolution.
\label{wz_halos.fig}}
\end{figure}

The solid line in Fig.~\ref{wz_halos.fig} shows the 95\% cl
constraint on $w(z)$ when this function is parameterized by $w_1$
and $w'$ for Case I above, with the choice of $z_1=0.35$ to
decorrelate these two parameters.  Two cases were considered,
each with a total of 10000 halos.  In the first, the objects were
binned into 8 redshift bins with $0.7<z<1.5$, as expected for the
DEEP sample~\cite{davis}.  In the second case, the objects were
binned in 15 redshift bins with $0<z<1.5$ (here $z_1=0.27$).
Filling in the low redshift end improves the constraint.

Finally, we show the constraint to $w(z)$ in the
case of 10000 halos with $0.7< z < 1.5$, 
but now assuming that there is
a 10\% (per bin) additional uncertainty due to the
evolution of the comoving number density of halos.
The constraint is now considerably weaker, and only
$w(z\approx 0.4)$ is determined accurately.  

\section{Optimal Strategies}\label{optimal-sec}

Here we consider strategies for the most
accurate determination of the cosmological parameters, $\Omega_M$,
$\Omega_X$ and the equation-of-state of
the dark energy, $w_X$, using high-redshift supernovae
(we add subscript 'X' to
distinguish it from the weight functions defined below). To this
end, we ask, given the cosmological parameters we want to
determine, what is the optimal redshfit distribution
to best constrain those parameters?

At first
glance this problem may appear of purely academic interest since
we are not free to put supernovae where we please.  However,
supernova observers have considerable freedom in choosing
redshift ranges for their searches, by using filters sensitive to
wavelengths corresponding to spectra at observed redshifts.
Moreover, supernovae are easier to discover than follow up,
and  the answer to the question we pose could well be
implemented in the choice of which supernovae are followed up.

In this section  we make three assumptions:

(i) Magnitude uncertainty, $\sigma_m$, is the same for each supernova
irrespective of redshift (this is a pretty good approximation
for the current data sets).

(ii)  Total number of supernovae observed is fixed (e.g., rather than
the total observing time).

(iii) The number of supernovae that can be found at any redshift
is not a limiting factor (this is not likely to be a serious
consideration).

(iv)  For simplicity we assume that type Ia supernovae are
standard candles; in fact, they are (at best) standardizable
candles whose peak luminosity is related to their rate of
decline brightness. 

None of these assumptions is required to use the formalism we
develop; rather, we make them for concreteness and simplicity.
Moreover, any or all of these assumptions can
be relaxed with the framework we present.  Finally, unless
the assumptions prove to be wildly wrong, the results will
not change much.

\subsection{Preliminaries}\label{sec-best}

We tackle the following problem: given $N$ supernovae and their
corresponding uncertainties, what distribution of these
supernovae in redshift would enable the most accurate
determination of $P$ cosmological parameters? In the case of more than
one parameter, we need to define what we mean by ``most accurate
determination''.  Since the uncertainty in measuring $P$
parameters simultaneously is described by an $P$-dimensional
ellipsoid (with the assumption that the total likelihood
function is Gaussian), we make a simple and, as it turns out,
mathematically tractable requirement that the ellipsoid have
minimal volume.  This corresponds to the best {\it local}
determination of the parameters.

The volume of the ellipsoid is given by
\begin{equation}
V \propto \det(F)^{-1/2},
\label{eq:vol_ellipsoid}
\end{equation}
where $F$ is the Fisher
matrix~\cite{Fisher-Jungman,Fisher-Tegmark}
\begin{equation}
F_{ij} = -\left< \partial^2 \ln L \over \partial p_i \partial p_j
 \right>_{\bf  y},
\end{equation}
and $L$ is the likelihood of observing data set ${\bf y}$ given
the parameters $p_1 \ldots p_P$. In Appendix \ref{app-A} we
present the derivation of Eq.~(\ref{eq:vol_ellipsoid}).
To minimize the volume of the uncertainty ellipsoid we
must maximize $\det(F)$.

\bigskip \bigskip
{\em Fisher matrix. }
The Fisher matrix for supernova
measurements was worked out in Ref.~\cite{CMB+SN};
we briefly review their results, with slightly
different notation and one important addition.

The supernova data consist of measurements of the
peak apparent magnitude of the individual supernovae,
$m_i$, which are related to the cosmological parameters by
\begin{equation}
m_i=5\log\;[H_0 d_L(z_n, \Omega_M, \Omega_\Lambda)]
+{\mathcal{M}}+\epsilon_i
\end{equation}
where $d_L$ is the luminosity distance to the supernova,
${\mathcal{M}}\equiv  M-5\log{H_0}+25$, $M$ is the
absolute magnitude of a type Ia supernova, and $\epsilon_i$ is the error
in the magnitude measurement (assumed to be Gaussian
with zero mean and standard deviation
$\sigma_m$). Note that ${\mathcal{M}}$ contains all dependence on
$H_0$, since $d_L\propto 1/H_0$.

The Fisher matrix is defined as~\cite{CMB+SN}

\begin{equation}
F_{ij}=\frac{1}{\sigma_m^2} \sum_{n=1}^{N} w_i(z_n)
w_j(z_n)
\end{equation}
where the $w$'s are weight functions given by
\begin{equation}
w_i(z) \equiv
{5\over\ln 10}
\left\{{\kappa S'[\kappa I(z)]\over S[\kappa I(z)]}
\left[{\partial I\over\partial p_i} - { I(z)\over  2\kappa^2}\right]
+ {1\over  2\kappa^2}
\right\},  \label{w_defined}
\end{equation}
if the parameter $p_i$ is $\Omega_M$ or $\Omega_X$, or else
\begin{equation}
w_i(z) \equiv
{5\over\ln 10}
\left[{\kappa S'[\kappa I(z)]\over S[\kappa I(z)]} \,
{\partial I\over\partial p_i}\right]
\end{equation}
if $p_i$ is $w_X$. Also

\begin{equation}
H_0\,d_L = (1+z)\, \frac{S(\kappa I)}{\kappa},
\end{equation}

\begin{equation}
S(x)=\left \{ \begin{array}{cl} \sinh(x), & \;\mbox{if}\;\; \Omega_0 > 1;\\
                                {x}      , & \;\mbox{if}\;\; \Omega_0 = 1;\\
 		                \sin(x) , & \;\mbox{if}\;\; \Omega_0 <  1.
              \end{array} \right.
\end{equation}
\vspace{-0.2cm}
\begin{eqnarray}
I(z, \Omega_M, \Omega_X, w_X) &=& \int_0^z H_0dx/H(x) \\[0.1cm]
\kappa^2  &=& 1- \Omega_M - \Omega_X.
\end{eqnarray}
When $w_X=-1$ (cosmological constant), we use $\Omega_\Lambda$
in place of $\Omega_X$.

In addition to $\Omega_M$, $\Omega_X$ and $w_X$, the
magnitude-redshift relation also includes the ``nuisance
parameter'' $\mathcal{M}$, which is a combination of the Hubble
parameter and absolute magnitude of supernovae, and which has to
be marginalized over in order to obtain constraints on the
parameters of interest. Ignoring $\mathcal{M}$ (that is, assuming
that $\mathcal{M} $ is known) leads to a 10\%-30\% underestimate
of the uncertainties in other parameters.  (Of course, accurate
knowledge of $H_0$ and a large local sample of supernovae could
be used to precisely determine $\mathcal{M}$ and eliminate this
additional parameter.)  For the moment we will ignore
$\mathcal{M}$ for clarity; later we will show that it is a simple
matter to include $\mathcal{M}$ as an additional parameter which
is marginalized over.

The Fisher matrix can be re-written as
\begin{eqnarray}
F_{ij} &=& {N\over\sigma_m^2}\int_0^{z_{\rm max}} g(z)  w_i(z)
w_j(z) dz, \label{F_ij}
\end{eqnarray}
where
\begin{equation}
g(z) = {1\over N}\sum_{n=1}^N \delta(z-z_n)
\label{define_g}
\end{equation}
is the (normalized) distribution of redshifts of the data and $z_{\rm
max}$ is the highest redshift probed in the survey. [$g(z)$ is
essentially a histogram of supernovae which is normalized to have unit
area.]  {\it Our goal
is to find $g(z)$ such that $\det(F)$ is maximal}.
Note that the maximization of $\det(F)$ will not depend on $N$
and $\sigma_m$, so we drop them for now. To consider
non-constant error $\sigma_m(z)$, one can simply absorb $\sigma_m(z)$
into the definition of weight functions $w(z)$.

\subsection{Results}\label{sec-results}

{\em One parameter.} As a warm-up, consider the case of measuring
a single cosmological parameter $p_1$.  We need to maximize
$\int_0^{z_{\rm max}} g(z)\, w_1^2(z)\,dz$, subject to $
\int_0^{z_{\rm max}} g(z) \, dz=1$ and $g(z)\geq 0$.  The
solution is a single delta function for $g(z)$ at the redshift
where $w_1(z)$ has a maximum. For {\it any} of our parameters,
$w_1(z)$ will have a maximum at $z_{\rm max}$.  This result is
hardly surprising: we have a one-parameter family of curves
$m(z)$, and the best way to distinguish between them is to have
all measurements at the redshift where the curves differ the
most, at $z_{\rm max}$.

\begin{figure}[ht]
\begin{center}
\includegraphics[width=3in, height=2.5in]{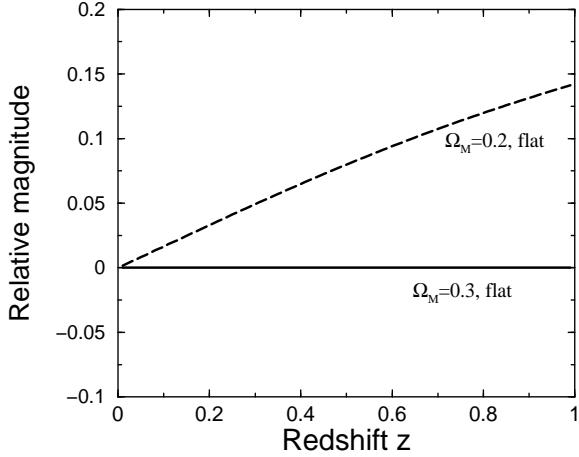}
\end{center}
\caption{Dependence of the magnitude-redshift relation upon
the single parameter $\Omega_M$,
relative to a flat Universe with $\Omega_M= 0.3$.
The maximum difference occurs at the highest redshift.}
\label{magdiff.fig}
\end{figure}

\begin{figure}[ht]
\includegraphics[width=3in, height=2.5in]{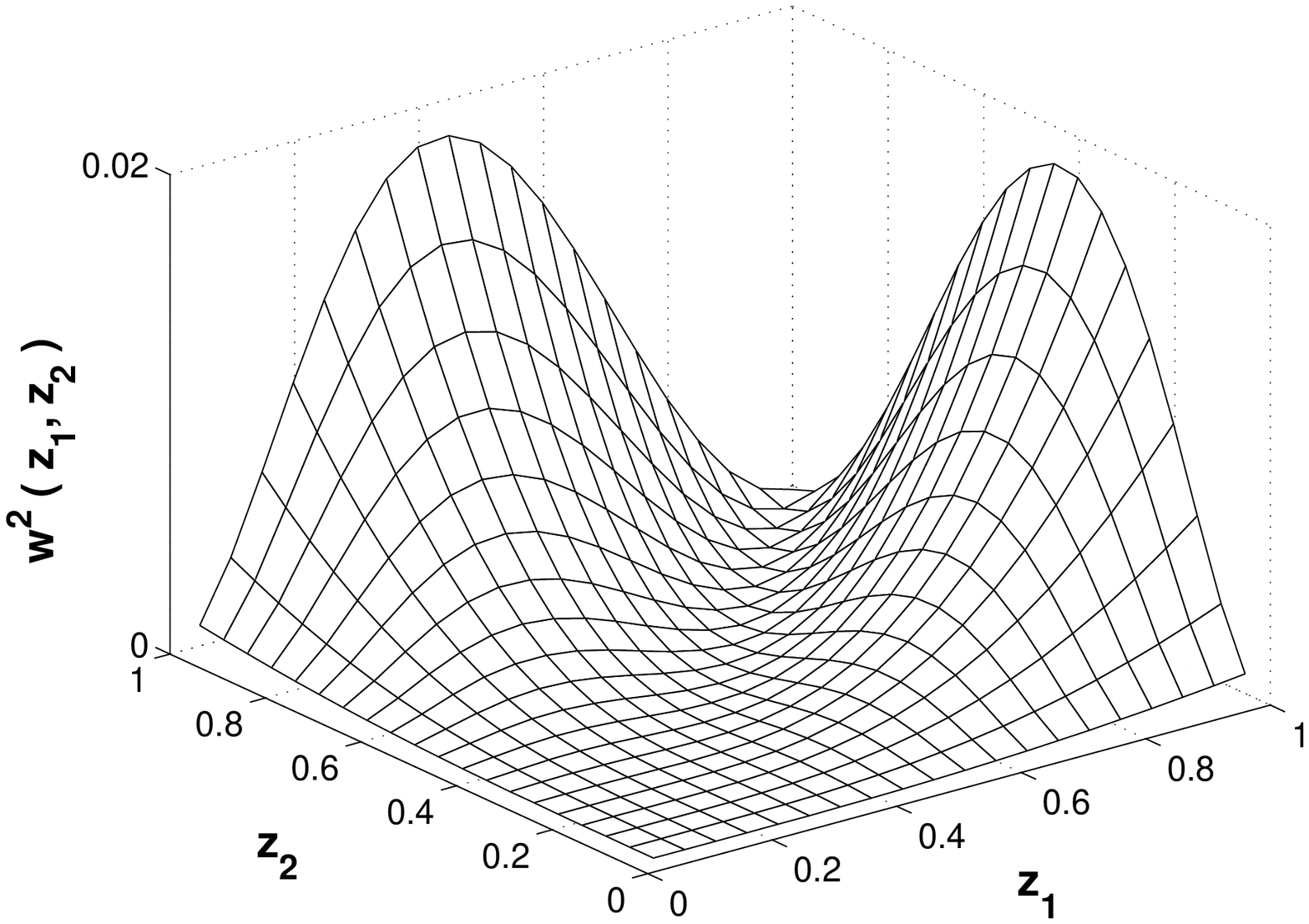}
\includegraphics[width=3in, height=2.5in]{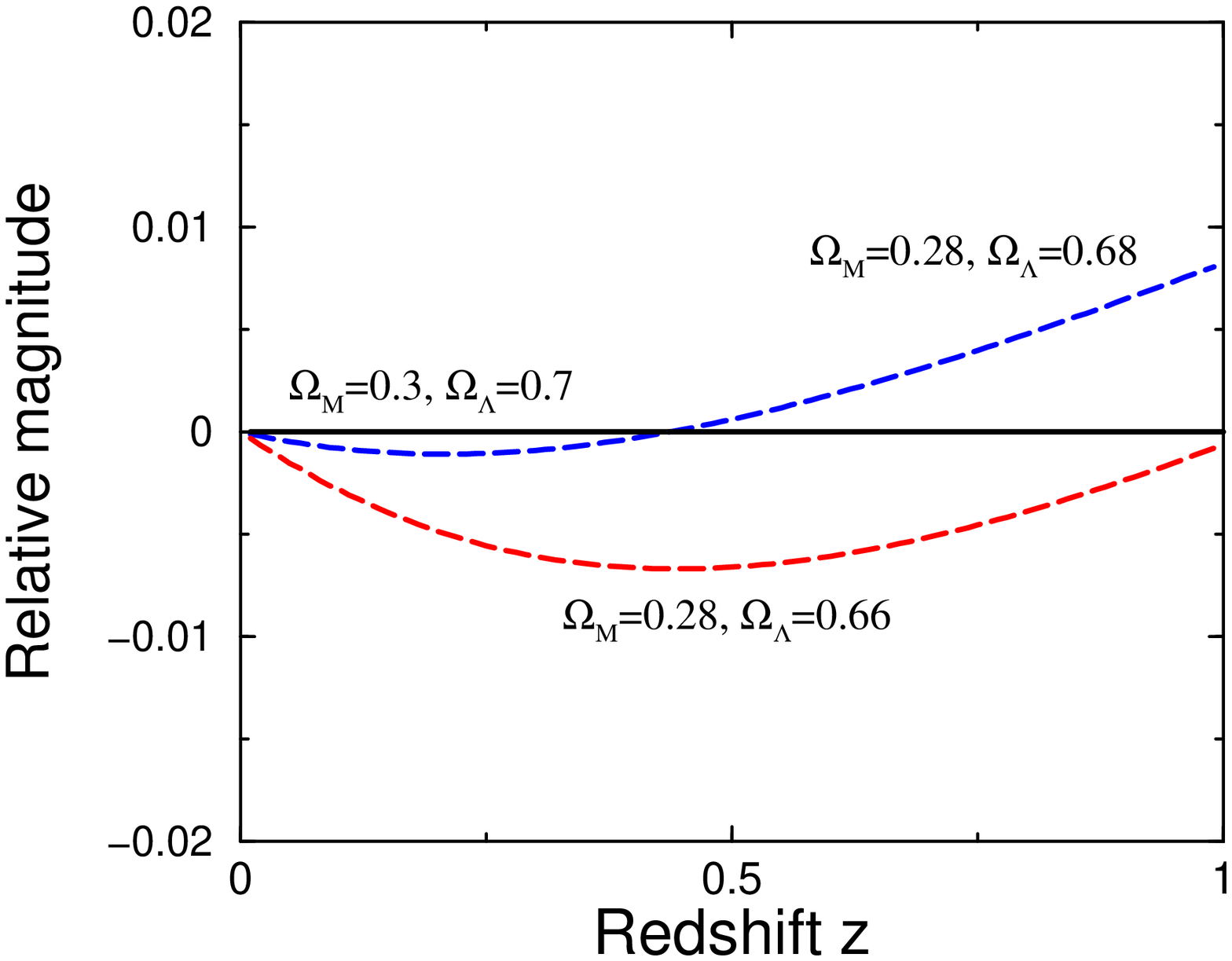}
\caption{{\it Top panel:} Function $w(z_1, z_2)^2$ for the case
when $\Omega_M=0.3$ and $\Omega_{\Lambda}=0.7$. {\it Bottom panel:}
Dependence of the magnitude-redshift relation upon
two parameter,  $\Omega_M$ and $\Omega_\Lambda$,
relative to a flat Universe with $\Omega_M= 0.3$.
Observations at more than one
redshift are needed to distinguish different models. }
\label{weight.fig}
\end{figure}

For example, Fig.~\ref{magdiff.fig} shows magnitude-redshift
curves for the fiducial $\Omega_M=0.3$ model with the assumption
$\Omega_{\Lambda}=1-\Omega_M$ (flat Universe). As $\Omega_M$ is
varied, the biggest difference in $m(z)$ is at the highest
redshift probed. In order to best constrain $\Omega_M$,
all supernovae should be located at $z_{\rm max} = 1.0$,

\bigskip\bigskip
{\em Two parameters.}
A more interesting -- and relevant -- problem is minimizing the
area of the error ellipse in the case of two
parameters, e.g., $\Omega_M$ and $w_X$ or $\Omega_M$
and $\Omega_X$.  The expression to maximize is now

\begin{eqnarray}
 && \int_{z=0}^{z_{\rm max}} g(z) \,w_1^2(z) dz
\int_{z=0}^{z_{\rm max}} g(z) \, w_2^2(z)  dz\;  - \nonumber \\[0.1cm]
& &\left (\int_{z=0}^{z_{\rm max}} g(z) \,w_1(z)\,  w_2(z) dz\right
)^2 \nonumber \\[0.2cm]
&=& {1\over 2} \int_{z_1=0}^{z_{\rm max}}\int_{z_2=0}^{z_{\rm
max}} g(z_1)\, g(z_2)
\,w(z_1, z_2)^2 \,dz_1\, dz_2, \label{area_analytic}
\end{eqnarray}
where $w(z_1, z_2)\equiv w_1(z_1)w_2(z_2) - w_1(z_2)w_2(z_1)$ is
a known function of redshifts and cosmological parameters (see
Fig.~\ref{weight.fig}) and $g(z)$ is subject to the same
constraints as before.

Despite the relatively harmless appearance of Eq.\
(\ref{area_analytic}), we found it impossible to maximize it
analytically. Fortunately, it is simple to find the
solution numerically. Returning to the discrete version
of Eq.~(\ref{F_ij}), we divide the interval $(0, z_{\rm
max})$ into $B$ bins with $g_i N$ supernovae in bin $i$. We
need to maximize
\begin{equation}
\sum_{i, j=1}^{B} g_i\,g_j\, w(z_i, z_j)^2 \label{g_ij}
 \end{equation}
subject to
\begin{equation}
\sum_{i=1}^{B} g_i=1 \,\,\,\,\, \mbox{and} \,\,\,\,\ g_i \geq 0.
\label{quadprog.constr}
\end{equation}

Equations (\ref{g_ij}) and (\ref{quadprog.constr}) define a quadratic
programming problem --- extremization of a quadratic function subject
to linear constraints. Since $w(z_1, z_2)^2$ is neither concave nor
convex (see Fig.~\ref{weight.fig}), we have to resort to brute force
maximization, and consider all possible values of $g_i$.
The result of this numerical maximization is that the
optimal distribution is two delta functions of equal magnitude:
\begin{equation}
g(z)= 0.50\; \delta(z-0.43) + 0.50 \; \delta(z-1.00) ,
\end{equation}
where all constants are accurate to $0.01$. Thus, half of the
supernovae should be at the highest available redshift, while the
other half at about 2/5 of the maximum redshift.

This result is not very sensitive to the maximum redshift probed, or
fiducial parameter values. If we increase the maximum available redshift to
$z_{\rm max}=1.5$, we find two delta functions of equal magnitude at
$z=0.57$ and $z=1.50$. If we change the fiducial values of
parameters to $\Omega_M=0.3$ and $\Omega_{\Lambda}=0$ (open
Universe), we find delta functions of equal magnitude at $z=0.47$
and $z=1.00$.

For a different choice for the two
parameters, $\Omega_M$ and $w_X$, with fiducial
values $\Omega_M=0.3$ and $w_X=-1$ and with the assumption of
flat Universe ($\Omega_X=1-\Omega_M$). we find a similar result
\begin{equation}
g(z)= 0.50 \;\delta(z-0.36) + 0.50\; \delta(z-1.00).
\end{equation}

\bigskip
{\em  Three or more parameters. }
We now consider parameter determination with three parameters
$\Omega_M$, $\Omega_X$ and $w_X$. Elements of the $3\times 3$ Fisher
matrix are calculated according to Eq.~(\ref{F_ij}), and we
again maximize $\det(F)$ as  described above. The result is
\begin{eqnarray}
g(z)&=& 0.33 \;\delta(z-0.21) + 0.34\; \delta(z-0.64) + \nonumber \\
 && 0.33\;\delta(z-1.00),
\end{eqnarray}
with all constants accurate to 0.01.  Hence we have three
delta functions of equal magnitude, with one of them at the
highest available redshift.

We have not succeeded in proving that $P>3$ parameters are best
measured if the redshift distribution is $P$ delta functions.
However, it is easy to prove that, {\em if} the data do
form $P$ delta functions, then those delta functions should be of
equal magnitude and their locations should be at coordinates
where the ``total'' weight function [e.g.\ $w(z_1, z_2)^2$ in case
of two parameters] has a global maximum (see Appendix
\ref{app-B}).  In practice, the number of cosmological
parameters to be determined from SNe Ia data is between one
and three, so considering more than three parameters is less
relevant.

\bigskip\bigskip {\em Marginalization over
$\mathcal{M}$. } So far we have been ignoring the parameter
$\mathcal{M}$, assuming that it is known (equivalently, that the
value of $H_0$ and the absolute magnitude of supernovae are
precisely known). This, of course, is not necessarily the case,
and $\mathcal{M}$ must be marginalized over to obtain
probabilities for the cosmological parameters. Fortunately, when
$\mathcal{M}$ is properly included, our results change in a
predictable and straightforward way.

Including $\mathcal{M}$ as an undetermined parameter, we now have
an ($P+1$)-dimensional ellipsoid ($P$ cosmological parameters plus
$\mathcal{M}$), and we want to minimize the volume of its
projection onto the $P$-dimensional space of cosmological
parameters. The equation of the $P$-dimensional projection
is

\begin{equation}
X^T F_{\rm proj} X=1\,.
\end{equation}
$F_{\rm proj}$ is obtained as follows: 1) Invert the original
$F$ to obtain the covariance matrix $F^{-1}$;\hspace{0.1cm} 2)
pick the desired $P$x$P$ subset of $F^{-1}$ and call it $F_{\rm
proj}^{-1}$; \hspace{0.08cm} 3) invert it to get $F_{\rm
proj}$.

Minimizing the volume of the projected ellipsoid we obtain the
result that the optimal supernova distribution is obtained with
$P$ delta functions in redshift obtained when ignoring
$\mathcal{M}$, plus a delta function at $z=0$. All $P+1$ delta
functions have the same magnitude. The explanation is simple:
the additional low redshift measurements pin down $\mathcal{M}$.

\bigskip\bigskip {\em Redshift dependent $\sigma_m$. }  The
optimal redshift distribution changes slightly if the uncertainty
in supernova measurements is redshift dependent. Suppose for
example that $\sigma_m = 0.15 + \sigma'z$, and that $z_{\rm max}=
2$.  In case of one parameter, the optimal location of SNe starts
changing from $z_{\rm max}=2$ only for $\sigma'>0.1$,
decreasing to $z=1.5$ for $\sigma'=0.2$. For the case of two or
more parameters, the optimal distribution is even more robust --
significant change occurs only for $\sigma'\gtrsim 0.3$ in the
case of two parameters, and only for $\sigma'\gtrsim 1$ in the
case of three.

\begin{figure}[!t]
\includegraphics[width=3in, height=2in]{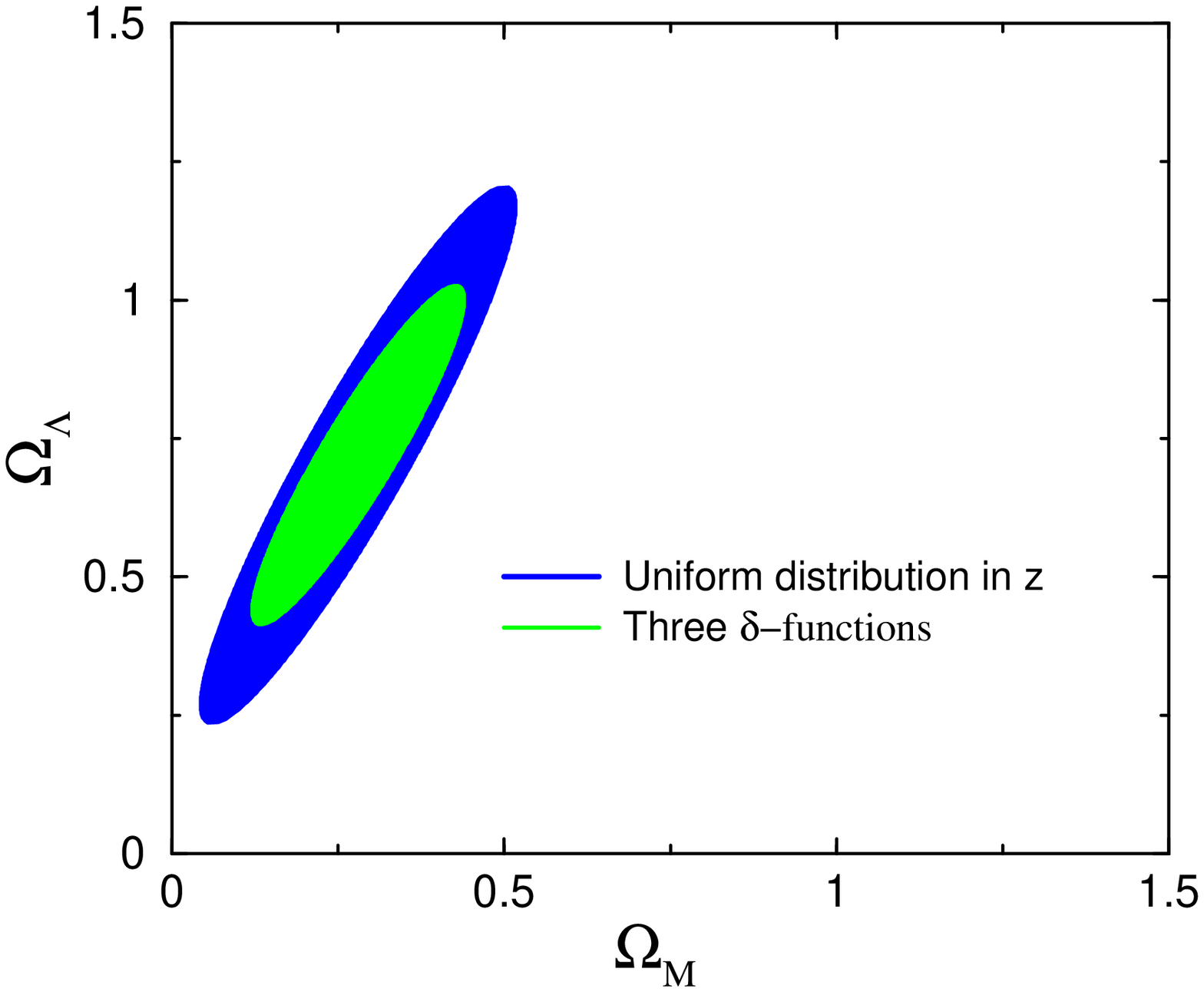}
\includegraphics[width=3in, height=2in]{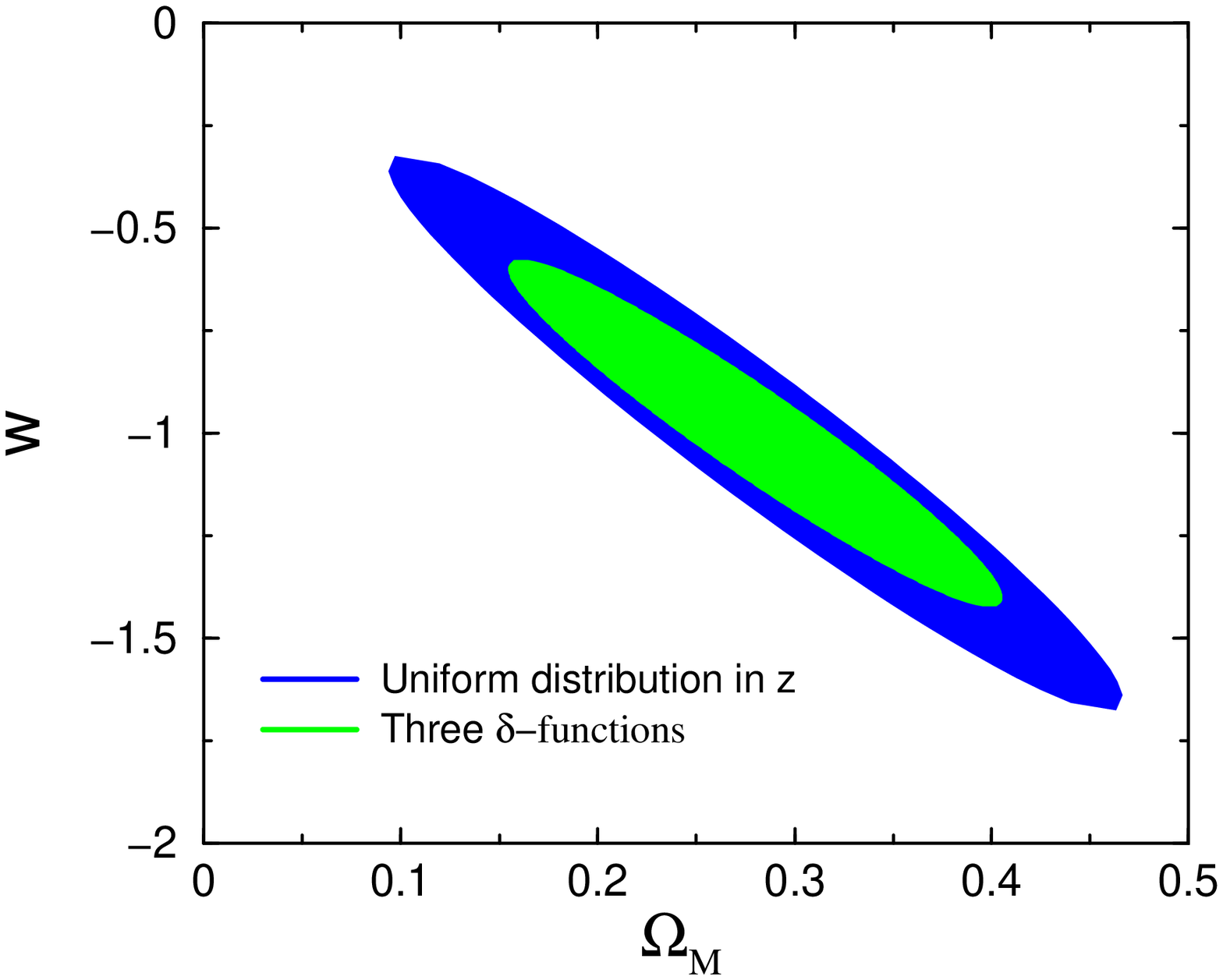}
\includegraphics[width=3in, height=2in]{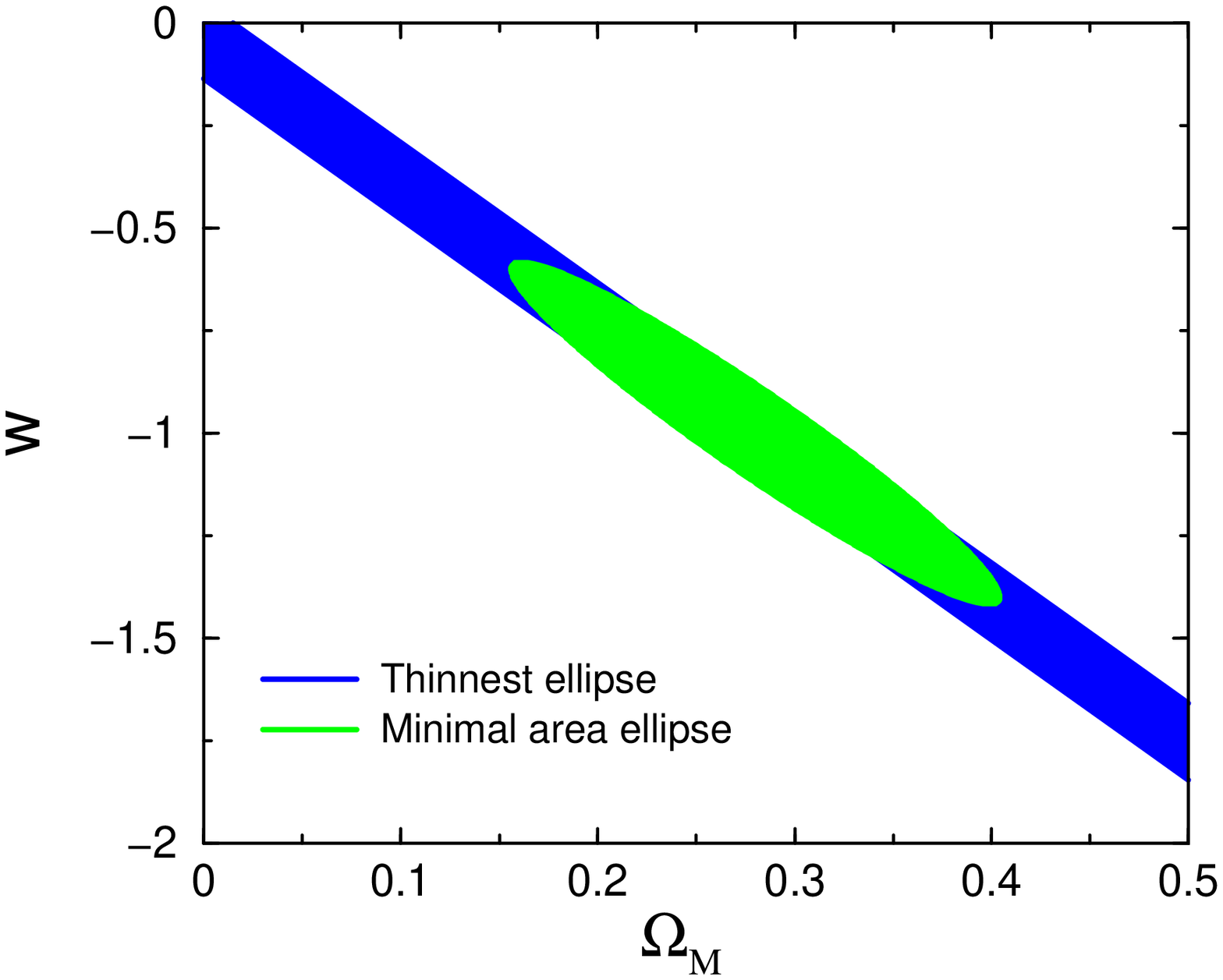}
\caption{{\em Top and middle panel:} Uniform (dark) vs.\ optimal
(light) distribution in redshift.  Shown are constraints on
$\Omega_M$ and $\Omega_{\Lambda}$ (top) and on $\Omega_M$ and $w$
for a flat Universe (middle) when $\mathcal{M}$ was marginalized
over.  For these results, 100 SNe were assumed with individual
uncertainties of $\sigma_m=0.15$ mag; the size of the error
ellipse scales as $\sigma_m/\sqrt{N}$.  {\em Bottom panel:}
Thinnest possible ellipse for given $N$ and $\sigma_m$ (dark) is
infinitely long in one direction. However, the smallest-area
ellipse (light) is almost as thin.}
\label{compare.ellipses.fig}
\end{figure}

\bigskip\bigskip {\em Optimal vs.\ uniform distribution.} Are the
advantages of the optimal distribution significant enough that
one should consider them seriously? In our opinion the answer is
yes, as we illustrate in the top panel of
Fig.~\ref{compare.ellipses.fig}.  This figure shows that the
area of the $\Omega_M$-$\Omega_{\Lambda}$ uncertainty ellipsoid
is more than two times smaller if the SNe have the
optimal distribution as opposed to
uniform distribution.  Similar results obtain for other
choices of the parameters.

\bigskip\bigskip
{\em  Thinnest ellipse. }
If we are using  SNe Ia alone to determine the cosmological
parameters, then we clearly want to minimize the area of the error
ellipse.  However, supernova
measurements will also be combined with other methods to determine
cosmological parameters.  A good example of the symbiosis is
combining CMB measurements with those of supernovae
\cite{Zaldarriaga,CMB+SN}. These methods together can improve the
determination of $\Omega_M$ and $\Omega_{\Lambda}$ by up to a factor of
10 as compared to either method alone by breaking the
degeneracy between the two parameters.  (The improvement is
largest when the error ellipses from the two methods are
comparable; in the case of SNAP and Planck, the projected
SNAP ellipse is so much smaller than Planck only improves
parameter determination by 5\% to 10\%; cf.,
Fig.~\ref{mat_w.SN_CMB.fig}.)

Finding the thinnest ellipse is a problem that we can solve using
our formalism.  Since the length of each axis of the ellipse is
proportional to the inverse square root of an eigenvalue of the
corresponding Fisher matrix, all we need to do is maximize the
larger eigenvalue of $F$ with respect to the distribution of the
supernovae $g(z)$.

The result is perhaps not surprising: to get the thinnest
ellipse, all supernova measurements should be at the same
(maximum) redshift, which leads to an infinitely
long ellipse.  We find that changing the supernovae redshift
distribution doesn't change the width of the
error ellipse greatly, but does change its length.
As a practical matter, we find the {\em smallest area} ellipse
is very close to being the thinnest ellipse (see
bottom panel of Fig.~\ref{compare.ellipses.fig}).

\bigskip\bigskip
{\em Reconstruction.}
In the spirit of our analyses above, we ask: what redshift distribution of
supernovae gives the smallest 95\% confidence region
for the reconstructed quintessence potential $V(\phi)$? To answer this
question, we perform a Monte-Carlo simulation by using different
distributions of supernovae and computing the average area of the
confidence region corresponding to each of them.

Uniform distribution of supernovae gives the best result among
the several distributions we put to test. This is not surprising,
because reconstruction of the potential consists in taking first
and second derivatives of the distance-redshift curve, and the
most accurate derivatives are obtained if the points are
distributed uniformly.  For comparison, Gaussian distribution of
supernovae with mean $\overline{z} =0.7$ and spread
$\sigma_z=0.4$ gives the area that is $10-20$\% larger.

\section{Conclusions}\label{conclude-sec}

Determining the nature of the dark energy that accounts for
2/3rds of the matter/energy in the Universe and is causing its
expansion to accelerate ranks as one of the most important
problems in both physics and astronomy.  At the moment, there is
very little theoretical guidance, and additional experimental
constraints are urgently needed. Because of its diffuse nature,
the effect of dark energy on the large-scale dynamics of the
Universe offers the most promising way to get this empirical
information.

The first step is to determine the average equation-of-state of
the dark energy.  CMB anisotropy, supernovae distance
measurements and number counts all appear promising.  The
Alcock-Paczynski shape test and the age of the Universe seem
somewhat less promising; the former because of the small size of
the effect (around 5\%); and the latter because the errors in the
two needed quantities, $H_0$ and $t_0$, are not likely to become
small enough in the near future.

The main sensitivity of the CMB to the dark energy is the $w$
dependence of the distance to the surface of last scattering,
which moves the positions of the acoustic peaks in the angular
power spectrum.  The CMB is much more sensitive to $\Omega_0$
than $w$, and the ultimate sensitivity of the CMB anisotropy to
$w$ will come from Planck, $\sigma_w\simeq 0.25$.

Probes of the low-redshift Universe (supernovae and number
counts) seem more promising.  In contrast to the CMB, they only
depend upon three cosmological parameters ($\Omega_M$, $\Omega_X$
and $w$), which will be effectively reduced to two ($\Omega_X$
and $w$) when precision CMB measurements determine $\Omega_0
=\Omega_M + \Omega_X$ to better than 1\%.  They are most
sensitive to $w$ between $z\sim 0.2$ and $z\sim 2$ (with
``sweet spot'' at $z\simeq 0.4$).

A high-quality sample of 2000 supernovae out to redshift $z\sim
1.5$ could determine $w$ to a precision of $\sigma_w =0.05$ (or
better if the optimal redshift distribution is achieved),
provided that the systematics associated with type Ia supernovae
can be controlled (e.g., luminosity evolution, photometric
errors, and dust).  A similar accuracy might be achieved by
number counts of galaxies out to $z\sim 1.5$ or of clusters of
galaxies.  The critical ingredient is understanding (or
independently measuring)  the evolution of the comoving number
density (in the case of galaxies to be better than 5\%).

More difficult, but very important, is a determination of, or
constraint to, the time variation of $w$.  If $w(z)$ is
parameterized to vary linearly (or logarithmically) with
redshift, and assuming perfect knowledge of $\Omega_M$ and
$\Omega_X$, a precision $\sigma_{w^\prime} \simeq 0.16$ might be
achieved by supernova  distance measurements ($w^\prime =
dw/dz$).  However, uncertainty in $\Omega_M$ significantly degrades
$\sigma_{w^\prime}$ (see Fig.~\ref{w_wprim.fig}).

Non-parametric reconstruction of either $w(z)$ or the potential-energy
curve for a quintessence model is the most demanding test, as it requires
the first and second derivatives of the luminosity distance $d_L$.
Even very accurate measurements of $d_L$ cannot constrain the small
bumps and wiggles which are crucial to reconstruction.  Without
some smoothing of the cosmological measurements, reconstruction is
impractical.  (The combination of number counts and supernova
measurements could determine $H(z)$ directly and eliminate the
dependence upon the second derivative of $d_L$.)

We have not addressed systematic error in any detail, and
for this reason our error forecasts could be very optimistic.  On
the other hand, the number of supernovae measured could be
larger and the uncertainties could be smaller than assumed (in
general, our error estimates scale as $\sigma_m/\sqrt{N}$).

We are at a very early stage in the study of dark energy.  Ways
of probing the dark energy not discussed here could well prove to
be equally or even more important.  Four examples come to mind.
First, the existence of a compelling model (or even one or
two-parameter class of models) would make the testing much
easier, as the predictions for $d_L(z)$ and other cosmological
observables could be directly compared to observations.  Second,
we have shown that one of the most powerful cosmological probes,
CMB anisotropy, has little leverage because dark energy was
unimportant at the time CMB anisotropies were formed ($z\sim
1100$).  Interesting ideas are now being discussed where the
ratio of dark energy to the total energy density does not decrease
dramatically with increasing redshift (or even stays roughly
constant)~\cite{manojetal,aaetal}; if correct, the power of the
CMB as a dark energy probe could be much greater.  Third, we have
assumed that the slight clumping of dark energy on large scales
is not an important probe.  While there are presently no models
where dark energy clumps significantly, if it does (or if the
clumping extends to smaller scales) CMB anisotropy and
large-scale structure measurements might have additional
leverage. Finally, it is possible that dark energy leads to other
observable effects such as a new long range force.

\bigskip

\begin{acknowledgments}
We would like to thank members of the Supernova Cosmology Project,
Daniel Eisenstein, Gil Holder, Wayne Hu, and Jeffrey Newman
for valuable discussions.  This work was supported by the DoE (at
Chicago and Fermilab) and by the NASA (at Fermilab by grant NAG 5-7092).
\end{acknowledgments}

\appendix
\section{proof of equation (38)}\label{app-A}

To derive Eq.\  (\ref{eq:vol_ellipsoid}), consider a general
uncertainty ellipsoid in $n$-dimensional parameter space.
The equation of this ellipsoid is
\begin{equation}
X^T F X =1,
\end{equation}
where $X=(x_1 x_2 \ldots x_P)$ is the vector of coordinates and
$F$ the Fisher matrix. Let us now choose coordinates so that the
ellipsoid has its axes parallel to the new coordinate axes.  Here
$X_{\rm rot}=UX$, where $U$ is the orthogonal matrix
corresponding to this rotation. The equation of the ellipsoid in
the new coordinate system is
\begin{equation}
X_{\rm rot}^T \,F_{\rm rot}\, X_{\rm  rot} =1,
\end{equation}
where $F_{\rm rot}=U F U^T$ is the Fisher matrix for the rotated
ellipsoid, and has the form $F_{\rm rot}=diag(1/\sigma_1^2,
\ldots, 1/\sigma_P^2)$. The volume of the ellipsoid is just

\begin{equation}
V\propto \prod_{i=1}^P \sigma_i = \det(F_{\rm rot})^{-1/2}.
\end{equation}

Then, since $\det(F)=\det(F_{\rm rot})$ and rotations preserve
volumes, we have

\begin{equation}
V \propto \det(F_{\rm rot})^{-1/2}= \det(F)^{-1/2}.
\end{equation}

This completes  the proof.

\section{Measuring $P$ parameters}\label{app-B}

We first prove the following: if we want to determine $P$
parameters by having measurements that form $P$ delta functions
in redshift (we have shown this is optimal for $P=1, 2, 3$ in
Sec.~\ref{sec-results}), then those delta functions should be
of equal magnitude.  The distribution of measurements is

\begin{equation}
g(z)= \sum_{i=1}^P \alpha_i \;\delta(z-z_i)
\end{equation}
with
\begin{equation}
 \sum_{i=1}^P \,\alpha_i =1. \label{app-constraint}
\end{equation}
We need to maximize
 \begin{equation}
\det F \propto \int  g(x_1) \ldots g(x_P)\,
W(x_1,\ldots x_P) \,dx_1\, \ldots dx_P  \label{app-W}
\end{equation}
where $x$'s are dummy variables and all integrations run from $0$
to $z_{\rm max}$. Here $W$ is given in terms of weight functions
$w_i$, and is symmetric under exchange of any two arguments and
zero if any two arguments are the same.  Then only one term in
the integrand of (\ref{app-W}) is non-zero and that expression
simplifies to
\begin{equation}
\det F \propto  \,P! \;\alpha_1\ldots \alpha_P\,
W(z_1,\ldots z_P). \label{app-prod-alphas}
\end{equation}
From (\ref{app-constraint}) and (\ref{app-prod-alphas}) one easily
shows that $\det F$ is maximized if $W$ has maximum at $(z_1,
\ldots, z_P)$ and
$\alpha_1=\alpha_2=\ldots=\alpha_P=1/P$. Therefore, the
delta functions must be of equal  magnitude, and located
where $W$ has its global maximum.

\end{document}